\documentclass[]{aa}  
\usepackage{natbib}
\bibpunct{(}{)}{;}{a}{}{,} % to follow the A&A style

\usepackage{graphicx}
\usepackage{txfonts}
\usepackage{hyperref}
\usepackage{lscape}
\usepackage{aalongtable}
\usepackage[svgnames]{xcolor}

\begin{document}

\title{Revisiting the correlation between stellar activity \\ and planetary surface gravity}

\author{P. Figueira\inst{1,2},
       M. Oshagh\inst{1,2,3},
	\,V. Zh. Adibekyan\inst{1,2},
	\and
	\,N. C. Santos\inst{1,2,3}}

   \institute{Centro de Astrof\'{i}sica, Universidade do Porto, Rua das Estrelas, 4150-762 Porto, Portugal\\
     \email{pedro.figueira@astro.up.pt}
     \and 
    Instituto de Astrof\' isica e Ci\^encias do Espa\c{c}o, Universidade do Porto, CAUP, Rua das Estrelas, PT4150-762 Porto, Portugal
    \and
    Departamento de F\'{i}sica e Astronomia, Faculdade de Ci\^{e}ncias, Universidade do Porto, Portugal
    }

   \date{}

% \abstract{}{}{}{}{} 
% 5 {} token are mandatory
 
  \abstract
  % context heading (optional)
  % {} leave it empty if necessary  
   {}
  % aims heading (mandatory)
   {We re-evaluate the correlation between planetary surface gravity and stellar host activity as measured by the index log($R'_{HK}$). This correlation, previously identified by Hartman (2010), is now analyzed in light of an extended measurements dataset, roughly 3 times larger than the original one.}
  % methods heading (mandatory)
   {We calculated the Spearman's rank correlation coefficient between the two quantities and its associated p-value. The correlation coefficient was calculated for both the full dataset and the star-planet pairs that follow the conditions proposed by Hartman (2010). In order to do so, we considered effective temperatures both as collected from the literature and from the SWEET-Cat catalog, which provides a more homogeneous and accurate effective temperature determination.}
  % results heading (mandatory)
   {The analysis delivers significant correlation coefficients, but with a lower value than those obtained by Hartman (2010). Yet, the two datasets are compatible, and we show that a correlation coefficient as large as previously published can arise naturally from a small-number statistics analysis of the current dataset. The correlation is recovered for star-planet pairs selected using the different conditions proposed by Hartman (2010). Remarkably, the usage of SWEET-Cat temperatures leads to larger correlation coefficient values. We highlight and discuss the role of the correlation betwen different parameters such as effective temperature and activity index. Several additional effects on top of those discussed previously were considered, but none fully explains the detected correlation.  In light of the complex issue discussed here, we encourage the different follow-up teams to publish their activity index values in the form of log($R'_{HK}$) index so that a comparison across stars and instruments can be pursued.}
  % conclusions heading (optional), leave it empty if necessary 
   {}

  \authorrunning{P. Figueira et al.}

  \maketitle
  \titlerunning{\thetitle}
%
%________________________________________________________________

\section{Introduction}\label{sec:Intro}
The search for exoplanets moves forward at a frenetic pace, and today we know more than 1700 planets in more than 1100 planetary systems. Several works have exploited the information gathered on exoplanet population, shedding some light on the properties of the population as whole \citep[e.g.][]{2011arXiv1103.2541H, 2011arXiv1109.2497M, 2012A&A...541A.139F, 2013A&A...549A.109B, 2013A&A...551A..90M, 2013ApJ...767...95D, 2014arXiv1406.6048S}. Interesting trends and correlations relating host and planetary parameters emerged \citep[see e.g.][]{2007ARA&A..45..397U}, and allowed us to better understand the mechanisms behind planetary formation and orbital evolution. However, some of the proposed trends are still waiting for a robust explanation. 

One of the most puzzling of the proposed correlations is the one between stellar activity and planetary surface gravity reported by \cite[][henceforth H10]{2010ApJ...717L.138H}. Starting from the data collected by \cite{2010ApJ...720.1569K}, H10 showed that for transiting extrasolar planets a statistically significant correlation existed between log($R'_{HK}$) and log(g$_p$), at a 99.5\% confidence level. This relationship is of particular interest because it can be connected with the ongoing debate on whether stellar activity can be related with the presence of exoplanets \citep[e.g.][]{2001MNRAS.325...55S, 2008ApJ...676..628S, 2011A&A...528A..58P, 2014A&A...565L...1P} and associated to exoplanet evaporation and evolution \cite[e.g.][]{2010A&A...514A..72L, 2012A&A...537L...3B}.

In this work we review the log($R'_{HK}$) - log(g$_p$) correlation by collecting the data on published exoplanets on log($R'_{HK}$) and stellar and planet properties, and re-evaluate the correlation in the light of an extended dataset. In Sect.\,\ref{sec:TheData} we describe how this new dataset was gathered and in Sect.\,\ref{sec:AR} we present our analysis and results. We discuss these results in Sect.\,\ref{sec:Discussion} and conclude on the subject in Sect.\,\ref{sec:Conclusions}.

%__________________________________________________________________

\section{Gathering the Data}\label{sec:TheData}

To collect the data we started from the sample of 39 stars of H10, who provide in their Table\,2 values for the log($R'_{HK}$) and log(g$_p$) of the planets orbiting them. Using the \textit{Exoplanet Encyclopaedia} \cite[][accessible from \url{exoplanet.eu}.]{2011A&A...532A..79S}, we searched the literature for transiting planets with mass and radius measurement orbiting stars with measured log($R'_{HK}$) values. We found a total of 69 new planet-star pairs\footnote{As of 07/07/2014.}, which when adding to those of H10 lead to a dataset 2.8 times larger than the original one. Out of these 69, 17 log($R'_{HK}$) were listed in the \textit{Exoplanet Orbit Database}\cite[][\url{exoplanet.org}]{2011PASP..123..412W}, while the others were collected from the literature as reported in the \textit{Exoplanet Encyclopaedia} website\footnote{We note that we could have started our analysis from the \textit{Exoplanet Orbit Database} but the number of planets and transiting planets listed on both sites showed that the \textit{Exoplanet Encyclopaedia} was more comprehensive.}.

The vast majority of works reported only one value of log($R'_{HK}$), obtained by co-adding spectra, or reported only the average or median value of the different spectra collected.\footnote{A notable exception was the recent work of \cite{2014ApJS..210...20M}, which provided the activity index values for each observation. In this case we used the average value as representative of each star.} Several works presented no error bars, and those who did so showed a large range of values, from 0.02 to 0.1 dex; for the sake of simplicity, and to avoid considering error bars only for some of our measurements, we refrained from using them. The planetary surface gravity was calculated from the planetary radius and mass, the later being corrected of the $\sin{i}$ factor. We note that for transiting planets (like those considered for this work) this correction is very small; for illustration the lowest inclination value of 82$^o$ analyzed corresponds to a correction factor of 0.99.

We would like to note that the planetary surface gravity can be determined directly from the transit and RV observables (such as the orbital eccentricity, semi-amplitude of RV, orbital period,and planet radius in unit of stellar radius), without any knowledge of the stellar mass value, as shown in \cite{2007MNRAS.379L..11S}. We obtained a very small difference (of the order of $\pm$0.005) between planetary surface gravity as calculated using both methods, which indicates that the correlation coefficient does not depend on the particular method used for the calculation of the surface gravity.

\begin{figure}

\includegraphics[width=9cm]{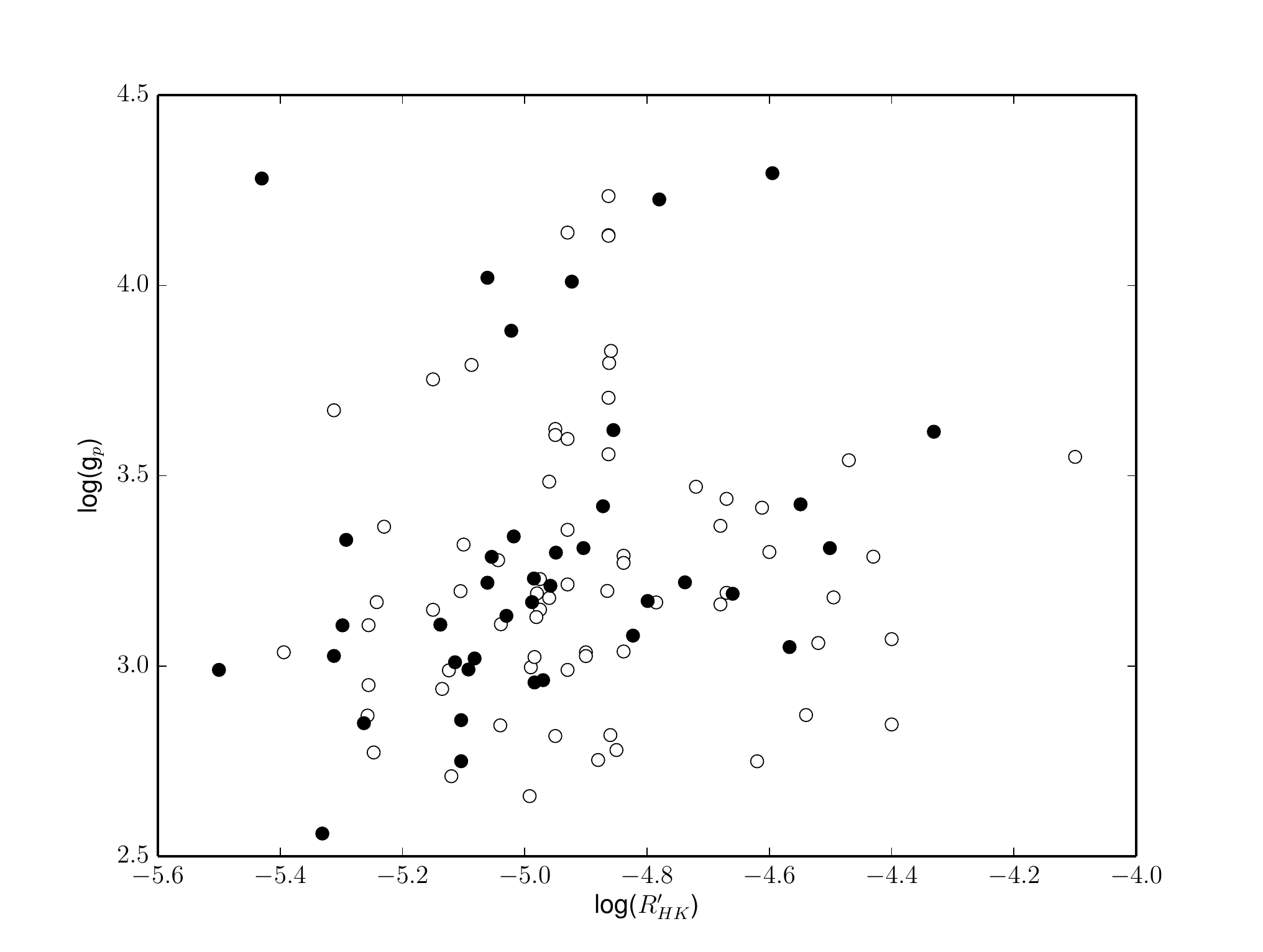}

\caption{Planetary gravity log(g$_p$) as a function of activity index log($R'_{HK}$) for the dataset studied in H10 (filled circled) and in this work (open circles).}\label{Fig:twodatasets}

\end{figure}

The collected values are presented in Appendix\,\ref{TableData} and the updated log($R'_{HK}$)-log(g$_p$) plot depicted in Fig.\,\ref{Fig:twodatasets}. 

The work of H10 considered different sub-datasets, by selecting stars and planets that fulfilled the following conditions:

\begin{enumerate}
 \item $M_{p}$\,$>$\,0.1\,$M_J$, $a$\,$<$\,0.1A.U., and 4200\,K\,$<$\,$T_{\mathrm{eff}}$\,$<$\,6200\,K;
 \item 4200\,K\,$<$\,$T_{\mathrm{eff}}$\,$<$\,6200\,K;
 \item No restrictions.
 
\end{enumerate}

in which $M_p$ is the mass of the planet, $a$ the semi-major axis of its planetary orbit and $T_{\mathrm{eff}}$ the effective temperature of the star. The constraint on the planetary parameters were used to select only massive close-by planets, and that on effective temperature to select only stars for which the log($R'_{HK}$) index had been calibrated \citep{1984ApJ...279..763N}. The evaluation of such conditions required the gathering of these quantities from the literature, a process done also using \textit{Exoplanet Encyclopaedia}. 

When compiling these values, we noticed the $T_{\mathrm{eff}}$ values could be different from those used by H10; importantly, the difference could be large enough to change the status of a planet-host star relative to the different conditions. So in order to use the most accurate $T_{\mathrm{eff}}$ measurements we reverted to the SWEET-Cat catalog \citep{2013A&A...556A.150S}, an updated catalog of stellar atmospheric parameters for all exoplanet-host stars. The stellar parameters were derived in an homogeneous way for 65\% of all planet-host and are dubbed ``baseline parameters''. For the remaining 35\% of the targets, the parameters were compiled from the literature (and whenever possible from uniform sources). It has been shown that the effective temperature $T_{\mathrm{eff}}$ derived by these authors is in very good agreement with that derived using other ``standard'' methods (e.g. the infrared flux method and interferometry), both for the low \citep{2013A&A...555A.150T} and high temperature range \citep{2008A&A...487..373S}. We note that two planets, WASP-69\,b and WASP 70\,b are still not listed in SWEET-Cat but had literature values of 4700 and 5700\,K respectively, allowing them to satisfy Conditions 1 and 2. As before, all the data gathered and their provenience is presented in Appendix\,\ref{TableData}.
%The $T_{\mathrm{eff}}$ for the stars considered in this work are depicted in Fig\,\ref{Fig:Temps}. 

\section{Analysis and Results}\label{sec:AR}

Our objective is to evaluate the correlation between the quantities log($R'_{HK}$) and log(g$_p$) and to do so we used the Spearman's rank correlation coefficient, as done by H10. For each correlation coefficient the p-value was calculated, i.e. the probability of having a larger or equal correlation coefficient under the hypothesis that the data pairs are uncorrelated (our null hypothesis). To do so we performed a simple yet robust non-parametric test: we shuffled the data pairs to create an equivalent uncorrelated dataset (using a Fisher-Yates shuffle to create unbiased datasets), and repeated the experiment 10 000 times. The correlation coefficient of the original dataset is compared with that of the shuffled population, and the original dataset z-score is calculated. The p-value is then calculated as the one-sided probability of having such a z-score from the observed Gaussian distribution \footnote{For more on this method and a python implementation of it the reader is referred to \cite{2013A&A...557A..93F, 2014A&A...566A..35S}.}.

\begin{figure*}

\includegraphics[width=16cm]{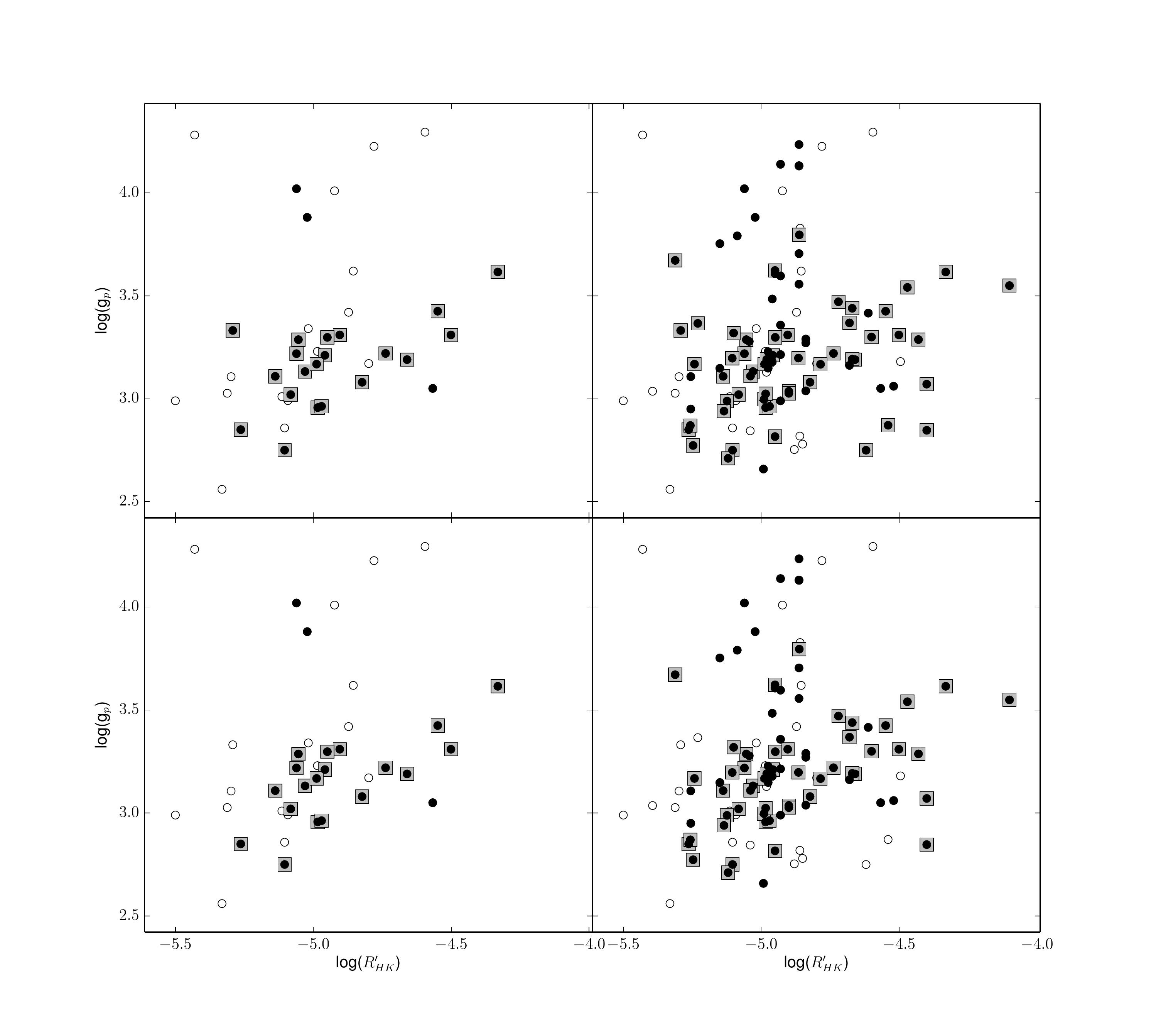}

\caption{Planetary gravity log(g$_p$) as a function of activity index log($R'_{HK}$) for the dataset studied in H10 only (\textit{left panels}) and the full one studied in this work (\textit{right panels}). The two rows represent the conditions when using $T_{\mathrm{eff}}$  from the literature ({\it upper panels}) and from SWEET-Cat ({\it lower panels}); the markers represent the three conditions considered: squares represent Condition 1, filled circles Condition 2 and open circles the whole dataset, i.e. Condition 3.}\label{Fig:2014_data}

\end{figure*}

The points selected when applied each of the conditions described in Sect.\ref{sec:TheData} and considering $T_{\mathrm{eff}}$ from the literature and SWEET-Cat are plotted in Fig.\,\ref{Fig:2014_data} for both the original H10 dataset and the new dataset. The results derived for each of the datasets and each of the conditions are presented in Table\,\ref{Table:Correlations}. We note that these values were not corrected for multiple testing\footnote{See, for instance, \url{http://en.wikipedia.org/wiki/Multiple_comparisons_problem}.}, and can be directly compared with those of H10. The correction for multiple testing would lead to an increase in p-value, an increase which depends on the estimated number of independent trials done and on the correction method used.

\begin{table*}

\caption{The number of points, Spearman's rank correlation coefficient, z-score, and p-value calculated as described in the text.} \label{Table:Correlations}

\centering
\begin{tabular}{cc|c|c|cc} \hline\hline
 \ \ dataset & Condition &  N & Spearman's rank & z-score & p-value [\%] \\ 
 \hline
Hartman data & 1 & 20 & 0.50 & 2.20  & 1.40 \\
Hartman data & 1 with SWEET-Cat $T_{\mathrm{eff}}$ & 19 & 0.69 & 2.92  & 0.18 \\

Hartman data & 2 &  23 & 0.29 & 1.34 & 9.04  \\
Hartman data & 2 with SWEET-Cat $T_{\mathrm{eff}}$ &  22 & 0.41 & 1.86 & 3.14  \\

Hartman data & 3 (whole dataset) & 39 & 0.45 & 2.76  & 0.28 \\

\hline
2014 data & 1 & 53 & 0.29 & 2.10 & 1.79 \\
2014 data & 1 with SWEET-Cat $T_{\mathrm{eff}}$ & 49 & 0.47 & 3.25 & 0.06 \\

2014 data & 2 &  84 & 0.21 & 1.88 & 3.01  \\
2014 data & 2 with SWEET-Cat $T_{\mathrm{eff}}$ &  80 & 0.30 & 2.68 & 0.37  \\

2014 data & 3 (whole dataset) & 108 & 0.26 & 2.70 & 0.35 \\

\hline 
\end{tabular}
\tablefoot{Note: the p-values were not corrected for multiple testing effect.}

\end{table*}

For the previously published H10 data, we could recover very similar values of correlation coefficient and p-value. However, when applying Condition 1 and 2, the correlation coefficient values depend on the source for the effective temperature constraint for the star. The correlation coefficient is the highest for Condition 1 and the lowest for Condition 2, for which p-values above 3\% cast serious doubt on the presence of the correlation. For the 2014 data, the correlation coefficients are always lower, in the range 0.21-0.47 in contrast with the previous values of 0.29-0.69. With p-values smaller than 1\%, the correlation seems significant, except for the application of Condition 1 and 2 using literature values for stellar effective temperature. 

In a nutshell, the analysis of data reveals that the correlations are in general still present in the extended 2014 dataset, even if with lower correlation coefficient values. However, these correlations are not significant for all the conditions tested, depending in particular on the choice of effective temperature constraint.

\section{Discussion}\label{sec:Discussion}

\subsection{Interpretation of the values and validity of the hypothesis}

\begin{figure}

\includegraphics[width=9cm]{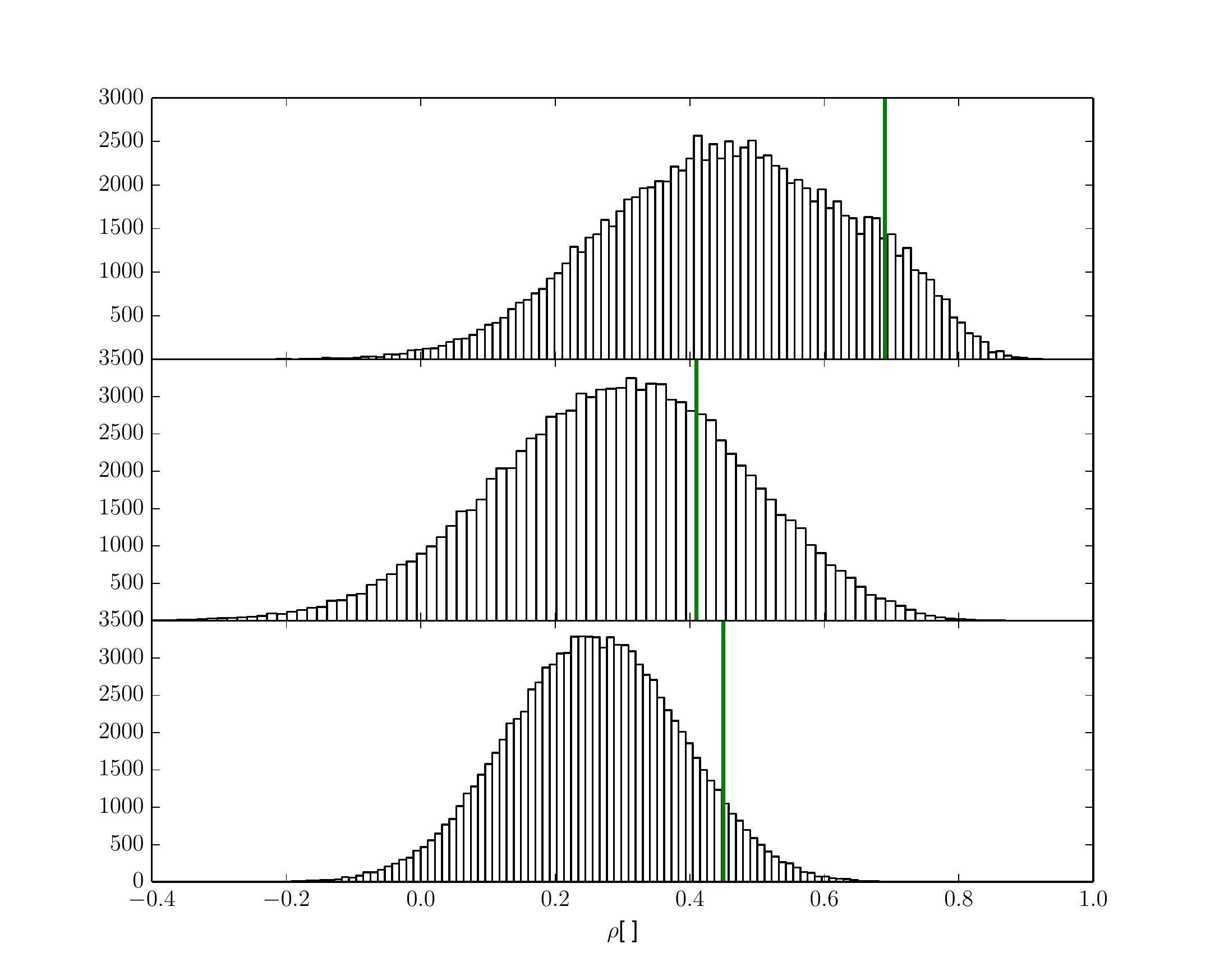}

\caption{Spearman's rank correlation coefficient for a mock population of log($R'_{HK}$)-log(g$_p$) pairs drawn from the 2014 data with the size of H10 dataset for the whole sample ({\it upper panel}), stars selected according to Condition 1 ({\it middle panel}), and stars selected according to Condition 2 ({\it lower panel}). The vertical lines represent the Spearman's Rank correlation coefficient values obtained using the set of stars present in H10.}\label{Fig:comp_dists}

\end{figure}

Before we dwell on the interpretation of the results obtained with our extended dataset, it is important to note that the correlation coefficient values obtained from it are systematically lower than those of H10. To understand to which extent these different values are compatible, we estimated the probability of obtaining the correlation coefficients of H10 by a chance draw from our current extended dataset. To do so, we selected from the 2014 dataset 39 random log($R'_{HK}$)-log(g$_p$) pairs\footnote{By ``random'' we stress that the pair chosen was random, but there was no shuffling or re-pairing of variables.}, the number of datapoints present in H10. We repeated this procedure 10 000 times and compared the distribution obtained with the value originally obtained by H10. After applying it to the whole dataset (i.e. Condition 3), we applied the same methodology for the stars selected using Condition 1 and Condition 2; using SWEET-Cat temperatures for this procedure, we selected 19 (or 22) out of the 49 (or 80) pairs available. The results are plotted in Fig.\,\ref{Fig:comp_dists}. The z-scores of the values using H10 data for the three cases of Condition 1, Condition 2 and Condition 3 (i.e. the whole dataset) were of 1.30, 0.27, and 1.52, and assuming a Gaussian distribution these correspond to a probabilities of 9.8, 39.5 and 6.4\% of having drawn an equal or larger value from the mock pairs distribution. The values obtained for the whole dataset and the stars selected using the different conditions were thus high when compared with the expected value from the distribution, but the associated probabilities were still large enough to consider it as a chance event. We note also that these probabilities do not correspond to independent events and their intersection is not the product of the three, and can be as high as the smallest of the three, 6.4\,\%.

Even though the analysis of H10 is thus compatible with the one presented here, the best estimation of the correlation coefficient is obtained with our extended dataset, with a value significantly lower than presented before. We note that the Spearman's rank correlation coefficient provides a measure for the concordance when ordering two variables, to which a p-value must be associated in order to understand how likely the correlation coefficient can arise from a chance event. The correlation coefficient corresponds to the Pearson's correlation coefficient of the ranked variables; squaring the correlation coefficient we obtain the coefficient of determination $r^2$, which we can equate to the explained or shared variance. So when we move from H10 dataset to the one presented here, the shared variance of the ranked variables decreases from 20\% to 7\%. However, it is important to recall that at this point we are referring to ranked variables, and it is hard to evaluate how this translates to our variable values prior to the ranking.

An important point to note from Table\,\ref{Table:Correlations} is the different number of stars selected when applying the different conditions. Since $T_{\mathrm{eff}}$  from different sources can vary appreciably, selecting a different $T_{\mathrm{eff}}$ origin for the Conditions can produce an appreciable difference in the number of stars selected; for instance for both HAT-P-32\,b and WASP-7, the difference is of 200\,K. As a consequence the correlation coefficients change when one uses temperature cut-offs from different sources. One can see that when applying the conditions to H10 data, the removal of a single point leads to different correlation coefficients: 0.50 vs. 0.69 for Condition 1, and 0.29 vs. 0.41 for Condition 2. Interestingly, for both datasets the largest values are obtained for Conditions based on SWEET-Cat temperatures. Performing the selection using a more accurate temperature determination should lead to a stronger correlation if one exists in the underlying data, and this can be seen as evidence for the presence of a correlation. However, this interpretation is complicated by the fact that log($R'_{HK}$) depends on the third power of the $(B-V)$ value of a star, which in turn depends on the $T_{\mathrm{eff}}$ \citep[see e.g.][]{2011A&A...526A..99S}. This might introduce non-obvious bias in the data, and the reader is cautioned about the limitation of such unguarded interpretations. 

A related issue is the validity of the $T_{\mathrm{eff}}$ boundaries chosen for the application of the two different criteria. The value of 4200\,K is strictly in line with the $R'_{HK}$ index calibration limits \citep{1984ApJ...279..763N}, but it might be over-optimistic given the difficulties at deriving precise parameters, especially for late K and early M stars \citep[e.g.][]{2012A&A...538A..25N, 2013A&A...555A.150T}. So we note that stars with temperature similar to the lower limit of the calibration are particularly prone to be misclassified. 

Attaching a single activity value to a star is a clear simplification of the issue, because the activity index value of a star varies throughout its activity cycle. As an example, Sun's activity index log($R'_{HK}$) ranges from a minimum of -5 to a maximum of -4.75  during its 11-years activity cycle \citep[e.g.][]{2011A&A...527A..82D}. Since many of the values reported for log($R'_{HK}$) in our study are obtained from a single observation, we caution that they might not be representative of a given star and a re-observation of these stars at different epochs for coverage of a full activity cycle might ultimately lead to different correlation coefficients. However, given the relatively low scatter expected to be introduced by it, this could not explain the measured index variability.
Along the same line, recent studies have demonstrated that a large number of stars have a higher activity level than that of the Sun. For instance, 67-75\% of the 150000 main sequence stars monitored by the {\it Kepler} satellite show a lower activity cycle than our host \citep{2013ApJ...769...37B}. In line with these studies, and as can be seen from Fig.\,\ref{Fig:2014_data}, most of the stars in our sample exhibit an activity level below -4.75: 34 of the 49 (69\,\%) selected using Condition1, 61 of the 80 (76\,\%) selected using Condition 2, and 85 of the 108 (79\,\%) selected using Condition 3 (i.e. the full dataset)\footnote{We note however that the detected planets are affected by selection bias, and their host properties might not reflect those of stars on the solar neighborhood.}. If one divides each of our samples using this activity cutoff, the lower-activity group still displays significant correlations while the group composed of more active stars does not, with p-values in excess of 30\,\%, as a consequence of small-number statistics.

One can also wonder to which extent can the radius be measured accurately for planets orbiting active stars. Some recent studies have shown that such determinations are bound to face some difficulties. The stellar spots which are non-occulted during the planetary transit can lead to an overestimation of the planet's radius determination of up to 3\% \cite{2009A&A...505.1277C}, while the occulted stellar spots during the transit can cause an underestimation on the planet's radius determination of up to 4\% \cite{2013A&A...556A..19O}.\footnote{Note that the upper limit on the underestimation/overestimation on the planet radius estimation was obtained for the case of the maximum Sunspot's filling factor, as measured during the maximum activity phase of the Solar cycle.}  This error on the planet' radius could lead to an error of 0.05 on the log(g$_p$) estimation. Since the range of values of log(g$_p$) is larger by more than one order of magnitude than the estimated error, we conclude that the sunspot's effect on the planetary density cannot be at the root of the correlation studied in this work.

\subsection{Biases of the data and correlation between other variables}

An important point already discussed in H10, is that of an observational bias of the sample. This hypothesis was discarded based on the fact that planetary mass and radius (among other parameters) showed a less significant correlation with the activity index. We re-evaluated this point by calculating the correlation between each of these parameters and log($R'_{HK}$) following the procedure described in Sect.\,\ref{sec:AR}. For each case we considered the complete dataset and the restricted datasets according to Condition 1 and 2. We present the corresponding results in Table\,\ref{Table:Correlations2}. The most significant correlation between the Mass and activity index occurs at 0.5\% and corresponds to Condition 1, while Radius and activity do not present a significant correlation for any of the conditions.

\begin{table}

\caption{Spearman's rank correlation coefficient, z-score, and p-value calculated between Mass/Radius/$T_{\mathrm{eff}}$ and log($R'_{HK}$), as described in the text.} \label{Table:Correlations2}

\centering
\begin{tabular}{ccc|cc} \hline\hline
 \ \  & Cond. &  Spearman's r. & z-score & p-value [\%] \\ 
 \hline
Mass& 1 & 0.37 & 2.55  & 0.53 \\

Mass & 2 & 0.11 & 0.96 & 16.8  \\

Mass & 3 & 0.06 & 0.61  & 27.1 \\

\hline
Radius & 1 & -0.03 & 0.20 & 42.0 \\

Radius & 2 & -0.04 & 0.38 & 35.0  \\

Radius & 3 & -0.14 & 1.42 & 7.84 \\

\hline
$T_{\mathrm{eff}}$ & 1 & -0.41 & 2.85 & 0.22 \\

$T_{\mathrm{eff}}$ & 2 & -0.51 & 4.51 & 0.00  \\

$T_{\mathrm{eff}}$ & 3 & -0.37 & 3.78 & 0.01 \\

\hline 
\end{tabular}
\tablefoot{Note: the Temperature criteria were set using SWEET-Cat; the p-values were not corrected for multiple testing effect.}

\end{table}

We tried to understand to which extent the measured correlations between Mass/Radius and the activity indicator could lead to the presented correlation with planetary surface gravity. With this aim we performed a simple test, which without being fully conclusive, is illustrative. We created mock distributions of 3 variables X, Y and Z which represent the activity index, the Mass, and the Radius, respectively. The activity index distribution X was represented by a standartized Gaussian distribution of N points, while the other two were created by selecting a slope and adding Gaussian noise such that the Spearman's rank between $\rho$(X,Y) and $\rho$(X,Z) delivered the value of our choice. By this procedure we created sets of three distributions with the same number of points N and correlation coefficients $\rho$(X,Y) and $\rho$(X,Z) as delivered in Table\,\ref{Table:Correlations2}. We then calculated $\rho$(X,Y.Z$^{-2}$), in which the latter distribution is analogous to that of the planetary surface gravity (modulus multiplicative constants, which do not have an impact on the correlation coefficient value). Repeating the experiment 10000 times, we concluded that the fraction of data sets with a correlation coefficient between our pseudo-surface gravity and pseudo-activity indicators higher than observed were of 1\% for Condition 1 and smaller than 0.1\% for conditions 2 and 3. Thus, it seems to be very unlikely to draw correlation coefficients as high as in Table\,\ref{Table:Correlations} starting from correlation coefficients between Mass/Radius and activity indicators as low as those in Table\,\ref{Table:Correlations2}. We stress that this test is more illustrative than definite, as there is a multitude of data pairs and distributions that deliver the same correlation coefficient. However, this simple quantitative analysis is well in line with the discussion of H10, and we have to conclude there is no reason to believe a correlation between the Mass and Radius and activity indicator could explain the high correlation value between planetary surface gravity and log($R'_{HK}$).

In addition to these previously explored correlations, we repeated the analysis to explore the one between the activity indicator and $T_{\mathrm{eff}}$. We did it using SWEET-Cat temperatures, and present the results in Table\,\ref{Table:Correlations2}. Very interestingly, effective temperature and log($R'_{HK}$) show a significant anti-correlation with value from 0.37 to 0.51, depending on the condition. Interpreting this correlation, and how it relates with the planetary gravity-activity correlation, is far more complicated. The activity indicator is, by construction, an instrument and effective temperature-independent ratio. The latter property is obtained by dividing the flux at the center of the Ca\,II line from that on the continuum, and correcting both line and continuum flux from their $(B-V)$ or effective temperature dependence. For more details the reader is referred to \cite{1984ApJ...279..763N}. However, a bias in the calculation as we move from instrument to instrument, or pipeline to pipeline, might introduce systematic biases in the activity indicator values, biases which will have an impact on the correlation studied here. For instance, the persistence of a dependence of log($R'_{HK}$) on $T_{\mathrm{eff}}$ might lead to a bias of the planetary parameters. In particular the well-known dependence of radius anomaly on effective temperature \citep[e.g.][]{2011ApJ...729L...7L} and the dependence of planetary mass on stellar mass via effective temperature \citep[e.g.][]{2007A&A...472..657L} could find a way into the studied correlation. In a similar way to the $T_{\mathrm{eff}}$, that is expected to be an absolute quantity and shows different results from author to author, the log($R'_{HK}$) index might show systematic differences that we unfortunately cannot explore by recalculating it in an homogeneous way, due to both the unavailability of most of the spectra and the gigantic task which is attempting an homogeneous reduction across such a range of instruments and spectra.

Previous works that evaluated the log($R'_{HK}$) distribution for a large number of stars reached different conclusions on the correlation found here. \cite{1996AJ....111..439H}, could not find a dependence of activity level on effective temperature (see their Fig.\,6 and associated discussion for details), while \cite{2011arXiv1107.5325L} presented opposite evidence from HARPS data. The latter authors noted that log($R'_{HK}$) values for G stars cluster around -5.0, while for K dwarfs they spread over the range from -4.7 to -5.0. The authors then noticed that this is in line with a slower decrease of activity with age as noted by \cite{2008ApJ...687.1264M}. The disagreement between studies raises  important questions; how can a large correlation be present in our sample while it is not for other samples is not obvious. This can be understood in two ways. First, the correlation between activity and effective temperature might be a consequence of the correlation between activity and planetary surface gravity. The existence of a dependency of activity on effective temperature would introduce a correlation with the latter. Alternatively, the causality might go in the other direction: the correlation between effective temperature and activity might introduce a correlation between surface gravity and activity. This would have to happen through another channel than only affecting the values of planetary Mass and planetary Radius of the sample; as we saw, the correlation between either of them and activity is very unlikely to accound for the high value of planetary surface gravity and activity index. An alternative would be an indirect effect of the planet on the activity level of the star, an effect which would depend on the effective temperature, for instance. As we see, this is an extremely involved issue and to fully address it one should dispose of a homogeneously derived activity indicator sample and preferably a control group of starts without planets.

The work of H10 had already discussed several possible explanations of the correlation found; these ranged from environmental effects acting on the planet, like extreme insolation and evaporation, or feed-back mechanisms into the star resulting from the proximity of the planet, like enhanced stellar activity. None of these can be robustly refuted, which makes the assessment of the existence of the correlation the more interesting. During the last stage of refereeing process we were informed of a contemporary theoretical work that proposes a new explanation for the correlation evaluated here. \cite{2014arXiv1410.8363L} argue that the correlation is due to the absorption by circumstellar material ejected by evaporating planets. Planets with lower atmospheric gravity have a greater mass loss being thus associated to a higher column density of circumstellar absorption, which in turn leads to a lower level of chromospheric emission as observed by us.

We would like to finish this section on a note to the ongoing existing transiting planet follow-up campaigns: we encourage the publication of log($R'_{HK}$) values instead of S index values. In this way the activity can be compared between stars and instruments, and more general trends like the discussed here can emerge. We also encourage the presentation of upper limits in the case of a non-detection of Ca\,II emission, so that a tobit-like analysis \citep[e.g.][]{2014A&A...570A..21F} can be performed.

\section{Conclusions}\label{sec:Conclusions}

We present an updated analysis on the correlation between stellar host activity as measured by the index log($R'_{HK}$) and orbiting planetary surface gravity. An updated dataset roughly 3 times larger than the original one shows significant correlation as delivered by Spearman's rank correlation coefficient, but with lower coefficient values. These values are compatible with those of H10 when considering small-number statistics on the current data, showing that the two datasets are not significantly different.

The correlation is recovered for star-planet pairs selected using the different conditions proposed by H10. Remarkably, the usage of the more homogeneous and accurate SWEET-Cat temperatures lead to larger correlation coefficient values. This can be interpreted as evidence for the existence of a real correlation in the data, but more complex dependencies might lurk within due to the dependence of the activity index on the effective temperature, of which we highlight importance and relation to the surface gravity-activity index correlation.

Several additional effects on top of those discussed by H10 were considered, but none fully explains the detected correlation. The very recent and contemporary work of \cite{2014arXiv1410.8363L} provides the best explanation to date. In light of the complex issue discussed here, we encourage the different follow-up teams to publish their activity index values in the form of log($R'_{HK}$) index so that a comparison across stars and instruments can be made.

%__________________________________________________________________

\begin{acknowledgements}
This work was supported by the European Research Council/European Community under the FP7 through Starting Grant agreement number 239953. PF and NCS acknowledge support by  Funda\c{c}\~ao para a Ci\^encia e a Tecnologia (FCT) through Investigador FCT contracts of reference IF/01037/2013 and IF/00169/2012, respectively, and POPH/FSE (EC) by FEDER funding through the program ``Programa Operacional de Factores de Competitividade - COMPETE''. 
VZhA and MO acknowledge the support from the Funda\c{c}\~ao para a Ci\^encia e Tecnologia, FCT (Portugal) in the form of the fellowships of reference SFRH/BPD/70574/2010 and SFRH/BD/51981/2012, respectively, from the FCT (Portugal). PF thanks Christophe Lovis and Maxime Marmier for a helpful discussion on the activity index definition, and Paulo Peixoto for the very hepful support on IT issues. We are indebted to the anonymous referee for constructive and insightful comments.
We warmly thank all those who develop the {\it Python} language and its scientific packages and keep them alive and free. This research has made use of the Exoplanet Orbit Database and the Exoplanet Data Explorer at exoplanets.org.

\end{acknowledgements}

% for the bibliography, at the end
\bibliographystyle{aa} % style aa.bst
\bibliography{Mybibliog,extra} % your references Yourfile.bib

\begin{thebibliography}{78}
\expandafter\ifx\csname natexlab\endcsname\relax\def\natexlab#1{#1}\fi

\bibitem[{{Albrecht} {et~al.}(2012){Albrecht}, {Winn}, {Butler}, {Crane},
  {Shectman}, {Thompson}, {Hirano}, \& {Wittenmyer}}]{2012ApJ...744..189A}
{Albrecht}, S., {Winn}, J.~N., {Butler}, R.~P., {et~al.} 2012, \apj, 744, 189

\bibitem[{{Anderson} {et~al.}(2013){Anderson}, {Collier Cameron}, {Delrez},
  {Doyle}, {Faedi}, {Fumel}, {Gillon}, {G{\'o}mez Maqueo Chew}, {Hellier},
  {Jehin}, {Lendl}, {Maxted}, {Pepe}, {Pollacco}, {Queloz}, {S{\'e}gransan},
  {Skillen}, {Smalley}, {Smith}, {Southworth}, {Triaud}, {Turner}, {Udry}, \&
  {West}}]{2013arXiv1310.5654A}
{Anderson}, D.~R., {Collier Cameron}, A., {Delrez}, L., {et~al.} 2013, ArXiv
  e-prints

\bibitem[{{Anderson} {et~al.}(2011){Anderson}, {Collier Cameron}, {Gillon},
  {Hellier}, {Jehin}, {Lendl}, {Queloz}, {Smalley}, {Triaud}, \&
  {Vanhuysse}}]{2011A&A...534A..16A}
{Anderson}, D.~R., {Collier Cameron}, A., {Gillon}, M., {et~al.} 2011, \aap,
  534, A16

\bibitem[{{Bakos} {et~al.}(2012){Bakos}, {Hartman}, {Torres}, {B{\'e}ky},
  {Latham}, {Buchhave}, {Csubry}, {Kov{\'a}cs}, {Bieryla}, {Quinn},
  {Szklen{\'a}r}, {Esquerdo}, {Shporer}, {Noyes}, {Fischer}, {Johnson},
  {Howard}, {Marcy}, {Sato}, {Penev}, {Everett}, {Sasselov}, {F{\H u}r{\'e}sz},
  {Stefanik}, {L{\'a}z{\'a}r}, {Papp}, \& {S{\'a}ri}}]{2012AJ....144...19B}
{Bakos}, G.~{\'A}., {Hartman}, J.~D., {Torres}, G., {et~al.} 2012, \aj, 144, 19

\bibitem[{{Ballard} {et~al.}(2011){Ballard}, {Fabrycky}, {Fressin},
  {Charbonneau}, {Desert}, {Torres}, {Marcy}, {Burke}, {Isaacson}, {Henze},
  {Steffen}, {Ciardi}, {Howell}, {Cochran}, {Endl}, {Bryson}, {Rowe}, {Holman},
  {Lissauer}, {Jenkins}, {Still}, {Ford}, {Christiansen}, {Middour}, {Haas},
  {Li}, {Hall}, {McCauliff}, {Batalha}, {Koch}, \&
  {Borucki}}]{2011ApJ...743..200B}
{Ballard}, S., {Fabrycky}, D., {Fressin}, F., {et~al.} 2011, \apj, 743, 200

\bibitem[{{Basri} {et~al.}(2013){Basri}, {Walkowicz}, \&
  {Reiners}}]{2013ApJ...769...37B}
{Basri}, G., {Walkowicz}, L.~M., \& {Reiners}, A. 2013, \apj, 769, 37

\bibitem[{{Beerer} {et~al.}(2011){Beerer}, {Knutson}, {Burrows}, {Fortney},
  {Agol}, {Charbonneau}, {Cowan}, {Deming}, {Desert}, {Langton}, {Laughlin},
  {Lewis}, \& {Showman}}]{2011ApJ...727...23B}
{Beerer}, I.~M., {Knutson}, H.~A., {Burrows}, A., {et~al.} 2011, \apj, 727, 23

\bibitem[{{B{\'e}ky} {et~al.}(2011){B{\'e}ky}, {Bakos}, {Hartman}, {Torres},
  {Latham}, {Jord{\'a}n}, {Arriagada}, {Bayliss}, {Kiss}, {Kov{\'a}cs},
  {Quinn}, {Marcy}, {Howard}, {Fischer}, {Johnson}, {Esquerdo}, {Noyes},
  {Buchhave}, {Sasselov}, {Stefanik}, {Perumpilly}, {L{\'a}z{\'a}r}, {Papp}, \&
  {S{\'a}ri}}]{2011ApJ...734..109B}
{B{\'e}ky}, B., {Bakos}, G.~{\'A}., {Hartman}, J., {et~al.} 2011, \apj, 734,
  109

\bibitem[{{Bonfils} {et~al.}(2013){Bonfils}, {Delfosse}, {Udry}, {Forveille},
  {Mayor}, {Perrier}, {Bouchy}, {Gillon}, {Lovis}, {Pepe}, {Queloz}, {Santos},
  {S{\'e}gransan}, \& {Bertaux}}]{2013A&A...549A.109B}
{Bonfils}, X., {Delfosse}, X., {Udry}, S., {et~al.} 2013, \aap, 549, A109

\bibitem[{{Borucki} {et~al.}(2013){Borucki}, {Agol}, {Fressin}, {Kaltenegger},
  {Rowe}, {Isaacson}, {Fischer}, {Batalha}, {Lissauer}, {Marcy}, {Fabrycky},
  {D{\'e}sert}, {Bryson}, {Barclay}, {Bastien}, {Boss}, {Brugamyer},
  {Buchhave}, {Burke}, {Caldwell}, {Carter}, {Charbonneau}, {Crepp},
  {Christensen-Dalsgaard}, {Christiansen}, {Ciardi}, {Cochran}, {DeVore},
  {Doyle}, {Dupree}, {Endl}, {Everett}, {Ford}, {Fortney}, {Gautier}, {Geary},
  {Gould}, {Haas}, {Henze}, {Howard}, {Howell}, {Huber}, {Jenkins}, {Kjeldsen},
  {Kolbl}, {Kolodziejczak}, {Latham}, {Lee}, {Lopez}, {Mullally}, {Orosz},
  {Prsa}, {Quintana}, {Sanchis-Ojeda}, {Sasselov}, {Seader}, {Shporer},
  {Steffen}, {Still}, {Tenenbaum}, {Thompson}, {Torres}, {Twicken}, {Welsh}, \&
  {Winn}}]{2013Sci...340..587B}
{Borucki}, W.~J., {Agol}, E., {Fressin}, F., {et~al.} 2013, Science, 340, 587

\bibitem[{{Borucki} {et~al.}(2011){Borucki}, {Koch}, {Basri}, {Batalha},
  {Brown}, {Bryson}, {Caldwell}, {Christensen-Dalsgaard}, {Cochran}, {DeVore},
  {Dunham}, {Gautier}, {Geary}, {Gilliland}, {Gould}, {Howell}, {Jenkins},
  {Latham}, {Lissauer}, {Marcy}, {Rowe}, {Sasselov}, {Boss}, {Charbonneau},
  {Ciardi}, {Doyle}, {Dupree}, {Ford}, {Fortney}, {Holman}, {Seager},
  {Steffen}, {Tarter}, {Welsh}, {Allen}, {Buchhave}, {Christiansen}, {Clarke},
  {Das}, {D{\'e}sert}, {Endl}, {Fabrycky}, {Fressin}, {Haas}, {Horch},
  {Howard}, {Isaacson}, {Kjeldsen}, {Kolodziejczak}, {Kulesa}, {Li}, {Lucas},
  {Machalek}, {McCarthy}, {MacQueen}, {Meibom}, {Miquel}, {Prsa}, {Quinn},
  {Quintana}, {Ragozzine}, {Sherry}, {Shporer}, {Tenenbaum}, {Torres},
  {Twicken}, {Van Cleve}, {Walkowicz}, {Witteborn}, \&
  {Still}}]{2011ApJ...736...19B}
{Borucki}, W.~J., {Koch}, D.~G., {Basri}, G., {et~al.} 2011, \apj, 736, 19

\bibitem[{{Bou{\'e}} {et~al.}(2012){Bou{\'e}}, {Figueira}, {Correia}, \&
  {Santos}}]{2012A&A...537L...3B}
{Bou{\'e}}, G., {Figueira}, P., {Correia}, A.~C.~M., \& {Santos}, N.~C. 2012,
  \aap, 537, L3

\bibitem[{{Brown} {et~al.}(2012){Brown}, {Cameron}, {Anderson}, {Enoch},
  {Hellier}, {Maxted}, {Miller}, {Pollacco}, {Queloz}, {Simpson}, {Smalley},
  {Triaud}, {Boisse}, {Bouchy}, {Gillon}, \&
  {H{\'e}brard}}]{2012MNRAS.423.1503B}
{Brown}, D.~J.~A., {Cameron}, A.~C., {Anderson}, D.~R., {et~al.} 2012, \mnras,
  423, 1503

\bibitem[{{Buchhave} {et~al.}(2010){Buchhave}, {Bakos}, {Hartman}, {Torres},
  {Kov{\'a}cs}, {Latham}, {Noyes}, {Esquerdo}, {Everett}, {Howard}, {Marcy},
  {Fischer}, {Johnson}, {Andersen}, {F{\H u}r{\'e}sz}, {Perumpilly},
  {Sasselov}, {Stefanik}, {B{\'e}ky}, {L{\'a}z{\'a}r}, {Papp}, \&
  {S{\'a}ri}}]{2010ApJ...720.1118B}
{Buchhave}, L.~A., {Bakos}, G.~{\'A}., {Hartman}, J.~D., {et~al.} 2010, \apj,
  720, 1118

\bibitem[{{Buchhave} {et~al.}(2011){Buchhave}, {Bakos}, {Hartman}, {Torres},
  {Latham}, {Andersen}, {Kov{\'a}cs}, {Noyes}, {Shporer}, {Esquerdo},
  {Fischer}, {Johnson}, {Marcy}, {Howard}, {B{\'e}ky}, {Sasselov}, {F{\H
  u}r{\'e}sz}, {Quinn}, {Stefanik}, {Szklen{\'a}r}, {Berlind}, {Calkins},
  {L{\'a}z{\'a}r}, {Papp}, \& {S{\'a}ri}}]{2011ApJ...733..116B}
{Buchhave}, L.~A., {Bakos}, G.~{\'A}., {Hartman}, J.~D., {et~al.} 2011, \apj,
  733, 116

\bibitem[{{Covino} {et~al.}(2013){Covino}, {Esposito}, {Barbieri}, {Mancini},
  {Nascimbeni}, {Claudi}, {Desidera}, {Gratton}, {Lanza}, {Sozzetti}, {Biazzo},
  {Affer}, {Gandolfi}, {Munari}, {Pagano}, {Bonomo}, {Collier Cameron},
  {H{\'e}brard}, {Maggio}, {Messina}, {Micela}, {Molinari}, {Pepe}, {Piotto},
  {Ribas}, {Santos}, {Southworth}, {Shkolnik}, {Triaud}, {Bedin}, {Benatti},
  {Boccato}, {Bonavita}, {Borsa}, {Borsato}, {Brown}, {Carolo}, {Ciceri},
  {Cosentino}, {Damasso}, {Faedi}, {Mart{\'{\i}}nez Fiorenzano}, {Latham},
  {Lovis}, {Mordasini}, {Nikolov}, {Poretti}, {Rainer}, {Rebolo L{\'o}pez},
  {Scandariato}, {Silvotti}, {Smareglia}, {Alcal{\'a}}, {Cunial}, {Di
  Fabrizio}, {Di Mauro}, {Giacobbe}, {Granata}, {Harutyunyan}, {Knapic},
  {Lattanzi}, {Leto}, {Lodato}, {Malavolta}, {Marzari}, {Molinaro},
  {Nardiello}, {Pedani}, {Prisinzano}, \& {Turrini}}]{2013A&A...554A..28C}
{Covino}, E., {Esposito}, M., {Barbieri}, M., {et~al.} 2013, \aap, 554, A28

\bibitem[{{Czesla} {et~al.}(2009){Czesla}, {Huber}, {Wolter}, {Schr{\"o}ter},
  \& {Schmitt}}]{2009A&A...505.1277C}
{Czesla}, S., {Huber}, K.~F., {Wolter}, U., {Schr{\"o}ter}, S., \& {Schmitt},
  J.~H.~M.~M. 2009, \aap, 505, 1277

\bibitem[{{Dressing} \& {Charbonneau}(2013)}]{2013ApJ...767...95D}
{Dressing}, C.~D. \& {Charbonneau}, D. 2013, \apj, 767, 95

\bibitem[{{Dumusque} {et~al.}(2014){Dumusque}, {Bonomo}, {Haywood},
  {Malavolta}, {Segransan}, {Buchhave}, {Collier Cameron}, {Latham},
  {Molinari}, {Pepe}, {Udry}, {Charbonneau}, {Cosentino}, {Dressing},
  {Figueira}, {Fiorenzano}, {Gettel}, {Harutyunyan}, {Horne}, {Lopez-Morales},
  {Lovis}, {Mayor}, {Micela}, {Motalebi}, {Nascimbeni}, {Phillips}, {Piotto},
  {Pollacco}, {Queloz}, {Rice}, {Sasselov}, {Sozzetti}, {Szentgyorgyi}, \&
  {Watson}}]{2014arXiv1405.7881D}
{Dumusque}, X., {Bonomo}, A.~S., {Haywood}, R.~D., {et~al.} 2014, ArXiv
  e-prints

\bibitem[{{Dumusque} {et~al.}(2011){Dumusque}, {Santos}, {Udry}, {Lovis}, \&
  {Bonfils}}]{2011A&A...527A..82D}
{Dumusque}, X., {Santos}, N.~C., {Udry}, S., {Lovis}, C., \& {Bonfils}, X.
  2011, \aap, 527, A82

\bibitem[{{Figueira} {et~al.}(2014){Figueira}, {Faria}, {Delgado-Mena},
  {Adibekyan}, {Sousa}, {Santos}, \& {Israelian}}]{2014A&A...570A..21F}
{Figueira}, P., {Faria}, J.~P., {Delgado-Mena}, E., {et~al.} 2014, \aap, 570,
  A21

\bibitem[{{Figueira} {et~al.}(2012){Figueira}, {Marmier}, {Bou{\'e}}, {Lovis},
  {Santos}, {Montalto}, {Udry}, {Pepe}, \& {Mayor}}]{2012A&A...541A.139F}
{Figueira}, P., {Marmier}, M., {Bou{\'e}}, G., {et~al.} 2012, \aap, 541, A139

\bibitem[{{Figueira} {et~al.}(2013){Figueira}, {Santos}, {Pepe}, {Lovis}, \&
  {Nardetto}}]{2013A&A...557A..93F}
{Figueira}, P., {Santos}, N.~C., {Pepe}, F., {Lovis}, C., \& {Nardetto}, N.
  2013, \aap, 557, A93

\bibitem[{{Gautier} {et~al.}(2012){Gautier}, {Charbonneau}, {Rowe}, {Marcy},
  {Isaacson}, {Torres}, {Fressin}, {Rogers}, {D{\'e}sert}, {Buchhave},
  {Latham}, {Quinn}, {Ciardi}, {Fabrycky}, {Ford}, {Gilliland}, {Walkowicz},
  {Bryson}, {Cochran}, {Endl}, {Fischer}, {Howell}, {Horch}, {Barclay},
  {Batalha}, {Borucki}, {Christiansen}, {Geary}, {Henze}, {Holman}, {Ibrahim},
  {Jenkins}, {Kinemuchi}, {Koch}, {Lissauer}, {Sanderfer}, {Sasselov},
  {Seager}, {Silverio}, {Smith}, {Still}, {Stumpe}, {Tenenbaum}, \& {Van
  Cleve}}]{2012ApJ...749...15G}
{Gautier}, III, T.~N., {Charbonneau}, D., {Rowe}, J.~F., {et~al.} 2012, \apj,
  749, 15

\bibitem[{{Gilliland} {et~al.}(2013){Gilliland}, {Marcy}, {Rowe}, {Rogers},
  {Torres}, {Fressin}, {Lopez}, {Buchhave}, {Christensen-Dalsgaard},
  {D{\'e}sert}, {Henze}, {Isaacson}, {Jenkins}, {Lissauer}, {Chaplin}, {Basu},
  {Metcalfe}, {Elsworth}, {Handberg}, {Hekker}, {Huber}, {Karoff}, {Kjeldsen},
  {Lund}, {Lundkvist}, {Miglio}, {Charbonneau}, {Ford}, {Fortney}, {Haas},
  {Howard}, {Howell}, {Ragozzine}, \& {Thompson}}]{2013ApJ...766...40G}
{Gilliland}, R.~L., {Marcy}, G.~W., {Rowe}, J.~F., {et~al.} 2013, \apj, 766, 40

\bibitem[{{Gillon} {et~al.}(2011){Gillon}, {Doyle}, {Lendl}, {Maxted},
  {Triaud}, {Anderson}, {Barros}, {Bento}, {Collier-Cameron}, {Enoch}, {Faedi},
  {Hellier}, {Jehin}, {Magain}, {Montalb{\'a}n}, {Pepe}, {Pollacco}, {Queloz},
  {Smalley}, {Segransan}, {Smith}, {Southworth}, {Udry}, {West}, \&
  {Wheatley}}]{2011A&A...533A..88G}
{Gillon}, M., {Doyle}, A.~P., {Lendl}, M., {et~al.} 2011, \aap, 533, A88

\bibitem[{{Hartman}(2010)}]{2010ApJ...717L.138H}
{Hartman}, J.~D. 2010, \apjl, 717, L138

\bibitem[{{Hartman} {et~al.}(2012){Hartman}, {Bakos}, {B{\'e}ky}, {Torres},
  {Latham}, {Csubry}, {Penev}, {Shporer}, {Fulton}, {Buchhave}, {Johnson},
  {Howard}, {Marcy}, {Fischer}, {Kov{\'a}cs}, {Noyes}, {Esquerdo}, {Everett},
  {Szklen{\'a}r}, {Quinn}, {Bieryla}, {Knox}, {Hinz}, {Sasselov}, {F{\H
  u}r{\'e}sz}, {Stefanik}, {L{\'a}z{\'a}r}, {Papp}, \&
  {S{\'a}ri}}]{2012AJ....144..139H}
{Hartman}, J.~D., {Bakos}, G.~{\'A}., {B{\'e}ky}, B., {et~al.} 2012, \aj, 144,
  139

\bibitem[{{Hartman} {et~al.}(2011{\natexlab{a}}){Hartman}, {Bakos}, {Kipping},
  {Torres}, {Kov{\'a}cs}, {Noyes}, {Latham}, {Howard}, {Fischer}, {Johnson},
  {Marcy}, {Isaacson}, {Quinn}, {Buchhave}, {B{\'e}ky}, {Sasselov}, {Stefanik},
  {Esquerdo}, {Everett}, {Perumpilly}, {L{\'a}z{\'a}r}, {Papp}, \&
  {S{\'a}ri}}]{2011ApJ...728..138H}
{Hartman}, J.~D., {Bakos}, G.~{\'A}., {Kipping}, D.~M., {et~al.}
  2011{\natexlab{a}}, \apj, 728, 138

\bibitem[{{Hartman} {et~al.}(2014){Hartman}, {Bakos}, {Torres}, {Kov{\'a}cs},
  {Johnson}, {Howard}, {Marcy}, {Latham}, {Bieryla}, {Buchhave}, {Bhatti},
  {B{\'e}ky}, {Csubry}, {Penev}, {de Val-Borro}, {Noyes}, {Fischer},
  {Esquerdo}, {Everett}, {Szklen{\'a}r}, {Zhou}, {Bayliss}, {Shporer},
  {Fulton}, {Sanchis-Ojeda}, {Falco}, {L{\'a}z{\'a}r}, {Papp}, \&
  {S{\'a}ri}}]{2014AJ....147..128H}
{Hartman}, J.~D., {Bakos}, G.~{\'A}., {Torres}, G., {et~al.} 2014, \aj, 147,
  128

\bibitem[{{Hartman} {et~al.}(2011{\natexlab{b}}){Hartman}, {Bakos}, {Torres},
  {Latham}, {Kov{\'a}cs}, {B{\'e}ky}, {Quinn}, {Mazeh}, {Shporer}, {Marcy},
  {Howard}, {Fischer}, {Johnson}, {Esquerdo}, {Noyes}, {Sasselov}, {Stefanik},
  {Fernandez}, {Szklen{\'a}r}, {L{\'a}z{\'a}r}, {Papp}, \&
  {S{\'a}ri}}]{2011ApJ...742...59H}
{Hartman}, J.~D., {Bakos}, G.~{\'A}., {Torres}, G., {et~al.}
  2011{\natexlab{b}}, \apj, 742, 59

\bibitem[{{H{\'e}brard} {et~al.}(2013){H{\'e}brard}, {Collier Cameron},
  {Brown}, {D{\'{\i}}az}, {Faedi}, {Smalley}, {Anderson}, {Armstrong},
  {Barros}, {Bento}, {Bouchy}, {Doyle}, {Enoch}, {G{\'o}mez Maqueo Chew},
  {H{\'e}brard}, {Hellier}, {Lendl}, {Lister}, {Maxted}, {McCormac}, {Moutou},
  {Pollacco}, {Queloz}, {Santerne}, {Skillen}, {Southworth}, {Tregloan-Reed},
  {Triaud}, {Udry}, {Vanhuysse}, {Watson}, {West}, \&
  {Wheatley}}]{2013A&A...549A.134H}
{H{\'e}brard}, G., {Collier Cameron}, A., {Brown}, D.~J.~A., {et~al.} 2013,
  \aap, 549, A134

\bibitem[{{Henry} {et~al.}(1996){Henry}, {Soderblom}, {Donahue}, \&
  {Baliunas}}]{1996AJ....111..439H}
{Henry}, T.~J., {Soderblom}, D.~R., {Donahue}, R.~A., \& {Baliunas}, S.~L.
  1996, \aj, 111, 439

\bibitem[{{Howard} {et~al.}(2012){Howard}, {Bakos}, {Hartman}, {Torres},
  {Shporer}, {Mazeh}, {Kov{\'a}cs}, {Latham}, {Noyes}, {Fischer}, {Johnson},
  {Marcy}, {Esquerdo}, {B{\'e}ky}, {Butler}, {Sasselov}, {Stefanik},
  {Perumpilly}, {L{\'a}z{\'a}r}, {Papp}, \& {S{\'a}ri}}]{2012ApJ...749..134H}
{Howard}, A.~W., {Bakos}, G.~{\'A}., {Hartman}, J., {et~al.} 2012, \apj, 749,
  134

\bibitem[{{Howard} {et~al.}(2011{\natexlab{a}}){Howard}, {Johnson}, {Marcy},
  {Fischer}, {Wright}, {Henry}, {Isaacson}, {Valenti}, {Anderson}, \&
  {Piskunov}}]{2011ApJ...730...10H}
{Howard}, A.~W., {Johnson}, J.~A., {Marcy}, G.~W., {et~al.} 2011{\natexlab{a}},
  \apj, 730, 10

\bibitem[{{Howard} {et~al.}(2011{\natexlab{b}}){Howard}, {Marcy}, {Bryson},
  {Jenkins}, {Rowe}, {Batalha}, {Borucki}, {Koch}, {Dunham}, {Gautier}, {Van
  Cleve}, {Cochran}, {Latham}, {Lissauer}, {Torres}, {Brown}, {Gilliland},
  {Buchhave}, {Caldwell}, {Christensen-Dalsgaard}, {Ciardi}, {Fressin}, {Haas},
  {Howell}, {Kjeldsen}, {Seager}, {Rogers}, {Sasselov}, {Steffen}, {Basri},
  {Charbonneau}, {Christiansen}, {Clarke}, {Dupree}, {Fabrycky}, {Fischer},
  {Ford}, {Fortney}, {Tarter}, {Girouard}, {Holman}, {Johnson}, {Klaus},
  {Machalek}, {Moorhead}, {Morehead}, {Ragozzine}, {Tenenbaum}, {Twicken},
  {Quinn}, {Isaacson}, {Shporer}, {Lucas}, {Walkowicz}, {Welsh}, {Boss},
  {Devore}, {Gould}, {Smith}, {Morris}, {Prsa}, \&
  {Morton}}]{2011arXiv1103.2541H}
{Howard}, A.~W., {Marcy}, G.~W., {Bryson}, S.~T., {et~al.} 2011{\natexlab{b}},
  ArXiv e-prints: 1103.2541

\bibitem[{{Kipping} {et~al.}(2011){Kipping}, {Hartman}, {Bakos}, {Torres},
  {Latham}, {Bayliss}, {Kiss}, {Sato}, {B{\'e}ky}, {Kov{\'a}cs}, {Quinn},
  {Buchhave}, {Andersen}, {Marcy}, {Howard}, {Fischer}, {Johnson}, {Noyes},
  {Sasselov}, {Stefanik}, {L{\'a}z{\'a}r}, {Papp}, {S{\'a}ri}, \& {F{\H
  u}r{\'e}sz}}]{2011AJ....142...95K}
{Kipping}, D.~M., {Hartman}, J., {Bakos}, G.~{\'A}., {et~al.} 2011, \aj, 142,
  95

\bibitem[{{Knutson} {et~al.}(2010){Knutson}, {Howard}, \&
  {Isaacson}}]{2010ApJ...720.1569K}
{Knutson}, H.~A., {Howard}, A.~W., \& {Isaacson}, H. 2010, \apj, 720, 1569

\bibitem[{{Kov{\'a}cs} {et~al.}(2010){Kov{\'a}cs}, {Bakos}, {Hartman},
  {Torres}, {Noyes}, {Latham}, {Howard}, {Fischer}, {Johnson}, {Marcy},
  {Isaacson}, {Sasselov}, {Stefanik}, {Esquerdo}, {Fernandez}, {L{\'a}z{\'a}r},
  {Papp}, \& {S{\'a}ri}}]{2010ApJ...724..866K}
{Kov{\'a}cs}, G., {Bakos}, G.~{\'A}., {Hartman}, J.~D., {et~al.} 2010, \apj,
  724, 866

\bibitem[{{Lanza}(2014)}]{2014arXiv1410.8363L}
{Lanza}, A.~F. 2014, ArXiv e-prints

\bibitem[{{Laughlin} {et~al.}(2011){Laughlin}, {Crismani}, \&
  {Adams}}]{2011ApJ...729L...7L}
{Laughlin}, G., {Crismani}, M., \& {Adams}, F.~C. 2011, \apjl, 729, L7

\bibitem[{{Lecavelier Des Etangs} {et~al.}(2010){Lecavelier Des Etangs},
  {Ehrenreich}, {Vidal-Madjar}, {Ballester}, {D{\'e}sert}, {Ferlet},
  {H{\'e}brard}, {Sing}, {Tchakoumegni}, \& {Udry}}]{2010A&A...514A..72L}
{Lecavelier Des Etangs}, A., {Ehrenreich}, D., {Vidal-Madjar}, A., {et~al.}
  2010, \aap, 514, A72

\bibitem[{{Lendl} {et~al.}(2012){Lendl}, {Anderson}, {Collier-Cameron},
  {Doyle}, {Gillon}, {Hellier}, {Jehin}, {Lister}, {Maxted}, {Pepe},
  {Pollacco}, {Queloz}, {Smalley}, {S{\'e}gransan}, {Smith}, {Triaud}, {Udry},
  {West}, \& {Wheatley}}]{2012A&A...544A..72L}
{Lendl}, M., {Anderson}, D.~R., {Collier-Cameron}, A., {et~al.} 2012, \aap,
  544, A72

\bibitem[{{Lendl} {et~al.}(2014){Lendl}, {Triaud}, {Anderson}, {Collier
  Cameron}, {Delrez}, {Doyle}, {Gillon}, {Hellier}, {Jehin}, {Maxted},
  {Neveu-VanMalle}, {Pepe}, {Pollacco}, {Queloz}, {S{\'e}gransan}, {Smalley},
  {Smith}, {Udry}, {Van Grootel}, \& {West}}]{2014A&A...568A..81L}
{Lendl}, M., {Triaud}, A.~H.~M.~J., {Anderson}, D.~R., {et~al.} 2014, \aap,
  568, A81

\bibitem[{{Lovis} {et~al.}(2011){Lovis}, {Dumusque}, {Santos}, {Bouchy},
  {Mayor}, {Pepe}, {Queloz}, {S{\'e}gransan}, \& {Udry}}]{2011arXiv1107.5325L}
{Lovis}, C., {Dumusque}, X., {Santos}, N.~C., {et~al.} 2011, ArXiv e-prints

\bibitem[{{Lovis} \& {Mayor}(2007)}]{2007A&A...472..657L}
{Lovis}, C. \& {Mayor}, M. 2007, \aap, 472, 657

\bibitem[{{Mamajek} \& {Hillenbrand}(2008)}]{2008ApJ...687.1264M}
{Mamajek}, E.~E. \& {Hillenbrand}, L.~A. 2008, \apj, 687, 1264

\bibitem[{{Mancini} {et~al.}(2014){Mancini}, {Southworth}, {Ciceri}, {Dominik},
  {Henning}, {J{\o}rgensen}, {Lanza}, {Rabus}, {Snodgrass}, {Vilela},
  {Alsubai}, {Bozza}, {Bramich}, {Calchi Novati}, {D'Ago}, {Figuera Jaimes},
  {Galianni}, {Gu}, {Harps{\o}e}, {Hinse}, {Hundertmark}, {Juncher}, {Kains},
  {Korhonen}, {Popovas}, {Rahvar}, {Skottfelt}, {Street}, {Surdej}, {Tsapras},
  {Wang}, \& {Wertz}}]{2014A&A...562A.126M}
{Mancini}, L., {Southworth}, J., {Ciceri}, S., {et~al.} 2014, \aap, 562, A126

\bibitem[{{Marcy} {et~al.}(2014){Marcy}, {Isaacson}, {Howard}, {Rowe},
  {Jenkins}, {Bryson}, {Latham}, {Howell}, {Gautier}, {Batalha}, {Rogers},
  {Ciardi}, {Fischer}, {Gilliland}, {Kjeldsen}, {Christensen-Dalsgaard},
  {Huber}, {Chaplin}, {Basu}, {Buchhave}, {Quinn}, {Borucki}, {Koch}, {Hunter},
  {Caldwell}, {Van Cleve}, {Kolbl}, {Weiss}, {Petigura}, {Seager}, {Morton},
  {Johnson}, {Ballard}, {Burke}, {Cochran}, {Endl}, {MacQueen}, {Everett},
  {Lissauer}, {Ford}, {Torres}, {Fressin}, {Brown}, {Steffen}, {Charbonneau},
  {Basri}, {Sasselov}, {Winn}, {Sanchis-Ojeda}, {Christiansen}, {Adams},
  {Henze}, {Dupree}, {Fabrycky}, {Fortney}, {Tarter}, {Holman}, {Tenenbaum},
  {Shporer}, {Lucas}, {Welsh}, {Orosz}, {Bedding}, {Campante}, {Davies},
  {Elsworth}, {Handberg}, {Hekker}, {Karoff}, {Kawaler}, {Lund}, {Lundkvist},
  {Metcalfe}, {Miglio}, {Silva Aguirre}, {Stello}, {White}, {Boss}, {Devore},
  {Gould}, {Prsa}, {Agol}, {Barclay}, {Coughlin}, {Brugamyer}, {Mullally},
  {Quintana}, {Still}, {Thompson}, {Morrison}, {Twicken}, {D{\'e}sert},
  {Carter}, {Crepp}, {H{\'e}brard}, {Santerne}, {Moutou}, {Sobeck}, {Hudgins},
  {Haas}, {Robertson}, {Lillo-Box}, \& {Barrado}}]{2014ApJS..210...20M}
{Marcy}, G.~W., {Isaacson}, H., {Howard}, A.~W., {et~al.} 2014, \apjs, 210, 20

\bibitem[{{Marmier} {et~al.}(2013){Marmier}, {S{\'e}gransan}, {Udry}, {Mayor},
  {Pepe}, {Queloz}, {Lovis}, {Naef}, {Santos}, {Alonso}, {Alves}, {Berthet},
  {Chazelas}, {Demory}, {Dumusque}, {Eggenberger}, {Figueira}, {Gillon},
  {Hagelberg}, {Lendl}, {Mardling}, {M{\'e}gevand}, {Neveu}, {Sahlmann},
  {Sosnowska}, {Tewes}, \& {Triaud}}]{2013A&A...551A..90M}
{Marmier}, M., {S{\'e}gransan}, D., {Udry}, S., {et~al.} 2013, \aap, 551, A90

\bibitem[{{Maxted} {et~al.}(2011){Maxted}, {Anderson}, {Collier Cameron},
  {Hellier}, {Queloz}, {Smalley}, {Street}, {Triaud}, {West}, {Gillon},
  {Lister}, {Pepe}, {Pollacco}, {S{\'e}gransan}, {Smith}, \&
  {Udry}}]{2011PASP..123..547M}
{Maxted}, P.~F.~L., {Anderson}, D.~R., {Collier Cameron}, A., {et~al.} 2011,
  \pasp, 123, 547

\bibitem[{{Mayor} {et~al.}(2011){Mayor}, {Marmier}, {Lovis}, {Udry},
  {S{\'e}gransan}, {Pepe}, {Benz}, {Bertaux}, {Bouchy}, {Dumusque}, {Lo Curto},
  {Mordasini}, {Queloz}, \& {Santos}}]{2011arXiv1109.2497M}
{Mayor}, M., {Marmier}, M., {Lovis}, C., {et~al.} 2011, ArXiv e-prints:
  1109.2497

\bibitem[{{McArthur} {et~al.}(2004){McArthur}, {Endl}, {Cochran}, {Benedict},
  {Fischer}, {Marcy}, {Butler}, {Naef}, {Mayor}, {Queloz}, {Udry}, \&
  {Harrison}}]{2004ApJ...614L..81M}
{McArthur}, B.~E., {Endl}, M., {Cochran}, W.~D., {et~al.} 2004, \apjl, 614, L81

\bibitem[{{Neves} {et~al.}(2012){Neves}, {Bonfils}, {Santos}, {Delfosse},
  {Forveille}, {Allard}, {Nat{\'a}rio}, {Fernandes}, \&
  {Udry}}]{2012A&A...538A..25N}
{Neves}, V., {Bonfils}, X., {Santos}, N.~C., {et~al.} 2012, \aap, 538, A25

\bibitem[{{Noyes} {et~al.}(1984){Noyes}, {Hartmann}, {Baliunas}, {Duncan}, \&
  {Vaughan}}]{1984ApJ...279..763N}
{Noyes}, R.~W., {Hartmann}, L.~W., {Baliunas}, S.~L., {Duncan}, D.~K., \&
  {Vaughan}, A.~H. 1984, \apj, 279, 763

\bibitem[{{O'Rourke} {et~al.}(2014){O'Rourke}, {Knutson}, {Zhao}, {Fortney},
  {Burrows}, {Agol}, {Deming}, {D{\'e}sert}, {Howard}, {Lewis}, {Showman}, \&
  {Todorov}}]{2014ApJ...781..109O}
{O'Rourke}, J.~G., {Knutson}, H.~A., {Zhao}, M., {et~al.} 2014, \apj, 781, 109

\bibitem[{{Oshagh} {et~al.}(2013){Oshagh}, {Santos}, {Boisse}, {Bou{\'e}},
  {Montalto}, {Dumusque}, \& {Haghighipour}}]{2013A&A...556A..19O}
{Oshagh}, M., {Santos}, N.~C., {Boisse}, I., {et~al.} 2013, \aap, 556, A19

\bibitem[{{Pepe} {et~al.}(2013){Pepe}, {Cameron}, {Latham}, {Molinari}, {Udry},
  {Bonomo}, {Buchhave}, {Charbonneau}, {Cosentino}, {Dressing}, {Dumusque},
  {Figueira}, {Fiorenzano}, {Gettel}, {Harutyunyan}, {Haywood}, {Horne},
  {Lopez-Morales}, {Lovis}, {Malavolta}, {Mayor}, {Micela}, {Motalebi},
  {Nascimbeni}, {Phillips}, {Piotto}, {Pollacco}, {Queloz}, {Rice}, {Sasselov},
  {S{\'e}gransan}, {Sozzetti}, {Szentgyorgyi}, \&
  {Watson}}]{2013Natur.503..377P}
{Pepe}, F., {Cameron}, A.~C., {Latham}, D.~W., {et~al.} 2013, \nat, 503, 377

\bibitem[{{Poppenhaeger} {et~al.}(2011){Poppenhaeger}, {Lenz}, {Reiners},
  {Schmitt}, \& {Shkolnik}}]{2011A&A...528A..58P}
{Poppenhaeger}, K., {Lenz}, L.~F., {Reiners}, A., {Schmitt}, J.~H.~M.~M., \&
  {Shkolnik}, E. 2011, \aap, 528, A58

\bibitem[{{Poppenhaeger} \& {Wolk}(2014)}]{2014A&A...565L...1P}
{Poppenhaeger}, K. \& {Wolk}, S.~J. 2014, \aap, 565, L1

\bibitem[{{Queloz} {et~al.}(2009){Queloz}, {Bouchy}, {Moutou}, {Hatzes},
  {H{\'e}brard}, {Alonso}, {Auvergne}, {Baglin}, {Barbieri}, {Barge}, {Benz},
  {Bord{\'e}}, {Deeg}, {Deleuil}, {Dvorak}, {Erikson}, {Ferraz Mello},
  {Fridlund}, {Gandolfi}, {Gillon}, {Guenther}, {Guillot}, {Jorda}, {Hartmann},
  {Lammer}, {L{\'e}ger}, {Llebaria}, {Lovis}, {Magain}, {Mayor}, {Mazeh},
  {Ollivier}, {P{\"a}tzold}, {Pepe}, {Rauer}, {Rouan}, {Schneider},
  {Segransan}, {Udry}, \& {Wuchterl}}]{2009A&A...506..303Q}
{Queloz}, D., {Bouchy}, F., {Moutou}, C., {et~al.} 2009, \aap, 506, 303

\bibitem[{{Quinn} {et~al.}(2012){Quinn}, {Bakos}, {Hartman}, {Torres},
  {Kov{\'a}cs}, {Latham}, {Noyes}, {Fischer}, {Johnson}, {Marcy}, {Howard},
  {Szentgyorgyi}, {F{\H u}r{\'e}sz}, {Buchhave}, {B{\'e}ky}, {Sasselov},
  {Stefanik}, {Perumpilly}, {Everett}, {L{\'a}z{\'a}r}, {Papp}, \&
  {S{\'a}ri}}]{2012ApJ...745...80Q}
{Quinn}, S.~N., {Bakos}, G.~{\'A}., {Hartman}, J., {et~al.} 2012, \apj, 745, 80

\bibitem[{{Saar} \& {Cuntz}(2001)}]{2001MNRAS.325...55S}
{Saar}, S.~H. \& {Cuntz}, M. 2001, \mnras, 325, 55

\bibitem[{{Santos} {et~al.}(2014){Santos}, {Mortier}, {Faria}, {Dumusque},
  {Adibekyan}, {Delgado-Mena}, {Figueira}, {Benamati}, {Boisse}, {Cunha},
  {Gomes da Silva}, {Lo Curto}, {Lovis}, {Martins}, {Mayor}, {Melo}, {Oshagh},
  {Pepe}, {Queloz}, {Santerne}, {S{\'e}gransan}, {Sozzetti}, {Sousa}, \&
  {Udry}}]{2014A&A...566A..35S}
{Santos}, N.~C., {Mortier}, A., {Faria}, J.~P., {et~al.} 2014, \aap, 566, A35

\bibitem[{{Santos} {et~al.}(2013){Santos}, {Sousa}, {Mortier}, {Neves},
  {Adibekyan}, {Tsantaki}, {Delgado Mena}, {Bonfils}, {Israelian}, {Mayor}, \&
  {Udry}}]{2013A&A...556A.150S}
{Santos}, N.~C., {Sousa}, S.~G., {Mortier}, A., {et~al.} 2013, \aap, 556, A150

\bibitem[{{Sato} {et~al.}(2012){Sato}, {Hartman}, {Bakos}, {B{\'e}ky},
  {Torres}, {Latham}, {Kov{\'a}cs}, {Csubry}, {Penev}, {Noyes}, {Buchhave},
  {Quinn}, {Everett}, {Esquerdo}, {Fischer}, {Howard}, {Johnson}, {Marcy},
  {Sasselov}, {Szklen{\'a}r}, {L{\'a}z{\'a}r}, {Papp}, \&
  {S{\'a}ri}}]{2012PASJ...64...97S}
{Sato}, B., {Hartman}, J.~D., {Bakos}, G.~{\'A}., {et~al.} 2012, \pasj, 64, 97

\bibitem[{{Schneider} {et~al.}(2011){Schneider}, {Dedieu}, {Le Sidaner},
  {Savalle}, \& {Zolotukhin}}]{2011A&A...532A..79S}
{Schneider}, J., {Dedieu}, C., {Le Sidaner}, P., {Savalle}, R., \&
  {Zolotukhin}, I. 2011, \aap, 532, A79

\bibitem[{{Shkolnik} {et~al.}(2008){Shkolnik}, {Bohlender}, {Walker}, \&
  {Collier Cameron}}]{2008ApJ...676..628S}
{Shkolnik}, E., {Bohlender}, D.~A., {Walker}, G.~A.~H., \& {Collier Cameron},
  A. 2008, \apj, 676, 628

\bibitem[{{Silburt} {et~al.}(2014){Silburt}, {Gaidos}, \&
  {Wu}}]{2014arXiv1406.6048S}
{Silburt}, A., {Gaidos}, E., \& {Wu}, Y. 2014, ArXiv e-prints

\bibitem[{{Sousa} {et~al.}(2011){Sousa}, {Santos}, {Israelian}, {Lovis},
  {Mayor}, {Silva}, \& {Udry}}]{2011A&A...526A..99S}
{Sousa}, S.~G., {Santos}, N.~C., {Israelian}, G., {et~al.} 2011, \aap, 526, A99

\bibitem[{{Sousa} {et~al.}(2008){Sousa}, {Santos}, {Mayor}, {Udry},
  {Casagrande}, {Israelian}, {Pepe}, {Queloz}, \&
  {Monteiro}}]{2008A&A...487..373S}
{Sousa}, S.~G., {Santos}, N.~C., {Mayor}, M., {et~al.} 2008, \aap, 487, 373

\bibitem[{{Southworth} {et~al.}(2007){Southworth}, {Wheatley}, \&
  {Sams}}]{2007MNRAS.379L..11S}
{Southworth}, J., {Wheatley}, P.~J., \& {Sams}, G. 2007, \mnras, 379, L11

\bibitem[{{Triaud} {et~al.}(2010){Triaud}, {Collier Cameron}, {Queloz},
  {Anderson}, {Gillon}, {Hebb}, {Hellier}, {Loeillet}, {Maxted}, {Mayor},
  {Pepe}, {Pollacco}, {S{\'e}gransan}, {Smalley}, {Udry}, {West}, \&
  {Wheatley}}]{2010A&A...524A..25T}
{Triaud}, A.~H.~M.~J., {Collier Cameron}, A., {Queloz}, D., {et~al.} 2010,
  \aap, 524, A25

\bibitem[{{Triaud} {et~al.}(2011){Triaud}, {Queloz}, {Hellier}, {Gillon},
  {Smalley}, {Hebb}, {Collier Cameron}, {Anderson}, {Boisse}, {H{\'e}brard},
  {Jehin}, {Lister}, {Lovis}, {Maxted}, {Pepe}, {Pollacco}, {S{\'e}gransan},
  {Simpson}, {Udry}, \& {West}}]{2011A&A...531A..24T}
{Triaud}, A.~H.~M.~J., {Queloz}, D., {Hellier}, C., {et~al.} 2011, \aap, 531,
  A24

\bibitem[{{Tsantaki} {et~al.}(2013){Tsantaki}, {Sousa}, {Adibekyan}, {Santos},
  {Mortier}, \& {Israelian}}]{2013A&A...555A.150T}
{Tsantaki}, M., {Sousa}, S.~G., {Adibekyan}, V.~Z., {et~al.} 2013, \aap, 555,
  A150

\bibitem[{{Udry} \& {Santos}(2007)}]{2007ARA&A..45..397U}
{Udry}, S. \& {Santos}, N.~C. 2007, \araa, 45, 397

\bibitem[{{Winn} {et~al.}(2011){Winn}, {Albrecht}, {Johnson}, {Torres},
  {Cochran}, {Marcy}, {Howard}, {Isaacson}, {Fischer}, {Doyle}, {Welsh},
  {Carter}, {Fabrycky}, {Ragozzine}, {Quinn}, {Shporer}, {Howell}, {Latham},
  {Orosz}, {Prsa}, {Slawson}, {Borucki}, {Koch}, {Barclay}, {Boss},
  {Christensen-Dalsgaard}, {Girouard}, {Jenkins}, {Klaus}, {Meibom}, {Morris},
  {Sasselov}, {Still}, \& {Van Cleve}}]{2011ApJ...741L...1W}
{Winn}, J.~N., {Albrecht}, S., {Johnson}, J.~A., {et~al.} 2011, \apjl, 741, L1

\bibitem[{{Wright} {et~al.}(2011){Wright}, {Fakhouri}, {Marcy}, {Han}, {Feng},
  {Johnson}, {Howard}, {Fischer}, {Valenti}, {Anderson}, \&
  {Piskunov}}]{2011PASP..123..412W}
{Wright}, J.~T., {Fakhouri}, O., {Marcy}, G.~W., {et~al.} 2011, \pasp, 123, 412

\end{thebibliography}

\Online

\begin{appendix}
%\begin{longtab}
%\begin{landscape}
 
\section{Literature Data}
In order to perform this study we gathered the data as described in Sect.\,\ref{sec:TheData}. The data employed is present in the Table below.
 
\begin{longtable}{ccccccccc}

\caption{The parameters for each star-planet pair considered in this study and their provenience.}\label{TableData}
\\
\hline\hline
planet &  log($R'_{HK}$)& log(g$_p$) & M[$M_{Jup}$] & R[$R_{Jup}$] & a[A.U.] & T$_{\mathrm{eff}}$[K]  & Cond. 1 & Cond. 2 \\ 
 \hline
 \endfirsthead
\caption{continued.}\\
\hline\hline
planet &  log($R'_{HK}$) & log(g$_p$) & M[$M_{Jup}$] & R[$R_{Jup}$] & a[A.U.] & T$_{\mathrm{eff}}$[K]  & Cond. 1 & Cond. 2 \\ 
\hline
\endhead
\hline
\endfoot
                                                             
CoRoT-1 b  &  -5.312   (1)  &  3.027  &  --  &  --  &  0.0254  &  6298 / 6397 (H) &  -- / --  &  -- / --  \\
CoRoT-2 b  &  -4.331   (1)  &  3.616  &  --  &  --  &  0.0281  &  5575 / 5697 (H) &  OK / OK  &  OK / OK  \\
GJ436 b  &  -5.298   (1)  &  3.107  &  --  &  --  &  0.02887  &  3684 / 3376 (H) &  -- / --  &  -- / --  \\
WASP-11-HAT-P-10 b  &  -4.823   (1)  &  3.080  &  --  &  --  &  0.0439  &  4980 / 4881 (H) &  OK / OK  &  OK / OK  \\
HAT-P-11 b  &  -4.567   (1)  &  3.050  &  --  &  --  &  0.053  &  4780 / 4624 (H) &  -- / --  &  OK / OK  \\
HAT-P-12 b  &  -5.104   (1)  &  2.750  &  --  &  --  &  0.0384  &  4650 / 4650  &  OK / OK  &  OK / OK  \\
HAT-P-13 b  &  -5.138   (1)  &  3.109  &  --  &  --  &  0.0426  &  5638 / 5653  &  OK / OK  &  OK / OK  \\
HAT-P-14 b  &  -4.855   (1)  &  3.620  &  --  &  --  &  0.0594  &  6600 / 6600  &  -- / --  &  -- / --  \\
HAT-P-1 b  &  -4.984   (1)  &  2.957  &  --  &  --  &  0.05561  &  5980 / 6076 (H) &  OK / OK  &  OK / OK  \\
HAT-P-2 b  &  -4.78   (1)  &  4.226  &  --  &  --  &  0.0674  &  6290 / 6290  &  -- / --  &  -- / --  \\
HAT-P-3 b  &  -4.904   (1)  &  3.310  &  --  &  --  &  0.03866  &  5224 / 5185  &  OK / OK  &  OK / OK  \\
HAT-P-4 b  &  -5.082   (1)  &  3.020  &  --  &  --  &  0.0446  &  5890 / 6054 (H) &  OK / OK  &  OK / OK  \\
HAT-P-5 b  &  -5.061   (1)  &  3.219  &  --  &  --  &  0.04079  &  5960 / 5960  &  OK / OK  &  OK / OK  \\
HAT-P-6 b  &  -4.799   (1)  &  3.171  &  --  &  --  &  0.05235  &  6570 / 6855 (H) &  -- / --  &  -- / --  \\
HAT-P-7 b  &  -5.018   (1)  &  3.341  &  --  &  --  &  0.0379  &  6259 / 6525 (H) &  -- / --  &  -- / --  \\
HAT-P-8 b  &  -4.985   (1)  &  3.230  &  --  &  --  &  0.0449  &  6200 / 6550 (H) &  -- / --  &  -- / --  \\
HAT-P-9 b  &  -5.092   (1)  &  2.991  &  --  &  --  &  0.053  &  6350 / 6350  &  -- / --  &  -- / --  \\
HD149026 b  &  -5.03   (1)  &  3.132  &  --  &  --  &  0.04288  &  6147 / 6162 (H) &  OK / OK  &  OK / OK  \\
HD17156 b  &  -5.022   (1)  &  3.881  &  --  &  --  &  0.1623  &  6079 / 6084 (H) &  -- / --  &  OK / OK  \\
HD189733 b  &  -4.501   (1)  &  3.310  &  --  &  --  &  0.03142  &  4980 / 5109 (H) &  OK / OK  &  OK / OK  \\
HD209458 b  &  -4.97   (1)  &  2.963  &  --  &  --  &  0.04747  &  6075 / 6118 (H) &  OK / OK  &  OK / OK  \\
HD80606 b  &  -5.061   (1)  &  4.020  &  --  &  --  &  0.449  &  5645 / 5574 (H) &  -- / --  &  OK / OK  \\
TrES-1  &  -4.738   (1)  &  3.220  &  --  &  --  &  0.0393  &  5230\tablefootmark{b} / 5226 (H) &  OK / OK  &  OK / OK  \\
TrES-2  &  -4.949   (1)  &  3.298  &  --  &  --  &  0.03556  &  5850 / 5795 (H) &  OK / OK  &  OK / OK  \\
TrES-3  &  -4.549   (1)  &  3.425  &  --  &  --  &  0.0226  &  5720 / 5502 (H) &  OK / OK  &  OK / OK  \\
TrES-4  &  -5.104   (1)  &  2.858  &  --  &  --  &  0.05084  &  6200 / 6293 (H) &  -- / --  &  -- / --  \\
WASP-12 b  &  -5.5   (1)  &  2.990  &  --  &  --  &  0.02293  &  6300 / 6313 (H) &  -- / --  &  -- / --  \\
WASP-13 b  &  -5.263   (1)  &  2.850  &  --  &  --  &  0.05379  &  5826 / 6025 (H) &  OK / OK  &  OK / OK  \\
WASP-14 b  &  -4.923   (1)  &  4.010  &  --  &  --  &  0.036  &  6475 / 6475  &  -- / --  &  -- / --  \\
WASP-17 b  &  -5.331   (1)  &  2.560  &  --  &  --  &  0.0515  &  6650 / 6794 (H) &  -- / --  &  -- / --  \\
WASP-18 b  &  -5.43   (1)  &  4.281  &  --  &  --  &  0.02047  &  6400 / 6526 (H) &  -- / --  &  -- / --  \\
WASP-19 b  &  -4.66   (1)  &  3.190  &  --  &  --  &  0.01616  &  5500 / 5591 (H) &  OK / OK  &  OK / OK  \\
WASP-1 b  &  -5.114   (1)  &  3.010  &  --  &  --  &  0.0382  &  6200 / 6252 (H) &  -- / --  &  -- / --  \\
WASP-2 b  &  -5.054   (1)  &  3.287  &  --  &  --  &  0.03138  &  5150 / 5109 (H) &  OK / OK  &  OK / OK  \\
WASP-3 b  &  -4.872   (1)  &  3.420  &  --  &  --  &  0.0313  &  6400 / 6448 (H) &  -- / --  &  -- / --  \\
XO-1 b  &  -4.958   (1)  &  3.211  &  --  &  --  &  0.0488  &  5750\tablefootmark{b} / 5754 (H) &  OK / OK  &  OK / OK  \\
XO-2 b  &  -4.988   (1)  &  3.168  &  --  &  --  &  0.0369  &  5340 / 5350 (H) &  OK / OK  &  OK / OK  \\
XO-3 b  &  -4.595   (1)  &  4.295  &  --  &  --  &  0.0454  &  6429 / 6429  &  -- / --  &  -- / --  \\
XO-4 b  &  -5.292   (1)  &  3.332  &  --  &  --  &  0.0555  &  5700 / 6397  &  OK / --  &  OK / --  \\

\hline
55Cnc e  &  -5.04349 (2\tablefootmark{a})  &  3.278  &  0.026  &  0.185  &  0.0156  &  5196 / 5279 (H) &  -- / --  &  OK / OK  \\
CoRoT-7 b  &  -4.612 (3)  &  3.416  &  0.023  &  0.148  &  0.0172  &  5313 / 5288 (H) &  -- / --  &  OK / OK  \\
HAT-P-15 b  &  -4.95 (4)  &  3.623  &  1.946  &  1.072  &  0.0964  &  5568 / 5568  &  OK / OK  &  OK / OK  \\
HAT-P-16 b  &  -4.862 (5)  &  3.796  &  4.193  &  1.289   &  0.0413  &  6158 / 6158  &  OK / OK  &  OK / OK  \\
HAT-P-17 b  &  -5.039 (6\tablefootmark{a})  &  3.110  &  0.530  &  1.01  &  0.0882  &  5246 / 5332 (H) &  OK / OK  &  OK / OK  \\
HAT-P-25 b  &  -4.99 (7\tablefootmark{a})  &  2.997  &  0.567  &  1.19  & 0.0466  &  5500 / 5500  &  OK / OK  &  OK / OK  \\
HAT-P-26 b  &  -4.992 (8\tablefootmark{a})  &  2.658  &  0.058  &  0.565   &  0.0479  &  5079 / 5011 (H) &  -- / --  &  OK / OK  \\
HAT-P-27-WASP-40 b  &  -4.785 (9)   &  3.167  &  0.662  &  1.055  &  0.0403  &  5300 / 5316 (H) &  OK / OK  &  OK / OK  \\
HAT-P-28 b  &  -4.984 (10)  &  3.024  &  0.626  &  1.212   &  0.0434  &  5680 / 5680  &  OK / OK  &  OK / OK  \\
HAT-P-29 b  &  -5.105 (10)  &  3.197  &  0.778  &  1.107   &  0.0667  &  6087 / 6087  &  OK / OK  &  OK / OK  \\
HAT-P-31 b  &  -5.312 (11)  &  3.672  &  2.173  &  1.07   &  0.055  &  6065 / 6065  &  OK / OK  &  OK / OK  \\
HAT-P-32 b  &  -4.62 (12)  &  2.750  &  0.941  &  2.037  &  0.0344  &  6001 / 6207  &  OK / --  &  OK / --  \\
HAT-P-33 b  &  -4.88 (12) &  2.753  &  0.764  &  1.827   &  0.0503  &  6401 / 6446  &  -- / --  &  -- / --  \\
HAT-P-34 b  &  -4.859 (13)  &  3.828  &  3.332  &  1.107 &  0.0677  &  6442 / 6442  &  -- / --  &  -- / --  \\
HAT-P-35 b  &  -5.242 (13)  &  3.168  &  1.055  &  1.332 &  0.0498  &  6096 / 6178 (H) &  OK / OK  &  OK / OK  \\
HAT-P-38 b  &  -5.124 (14\tablefootmark{a})  &  2.989  &  0.267  &  0.825&  0.0523  &  5330 / 5330  &  OK / OK  &  OK / OK  \\
HAT-P-39 b  &  -4.85 (15\tablefootmark{a})  &  2.779  &  0.600  &  1.571   &  0.0509  &  6430 / 6340  &  -- / --  &  -- / --  \\
HAT-P-40 b  &  -5.12 (15\tablefootmark{a})  &  2.711  &  0.620  &  1.73   &  0.0608  &  6080 / 6080  &  OK / OK  &  OK / OK  \\
HAT-P-41 b  &  -5.04 (15\tablefootmark{a})  &  2.844  &  0.800  &  1.685   &  0.0426  &  6390 / 6390  &  -- / --  &  -- / --  \\
HAT-P-44 b  &  -5.247 (16)  &  2.773  &  0.392  &  1.28  &  0.0507  &  5295 / 5295  &  OK / OK  &  OK / OK  \\
HAT-P-45 b  &  -5.394 (16)  &  3.036  &  0.892  &  1.426  &  0.0452  &  6330 / 6330  &  -- / --  &  -- / --  \\
HAT-P-46 b  &  -5.257 (16)  &  2.870  &  0.494  &  1.284  &  0.0577  &  6120 / 6120  &  OK / OK  &  OK / OK  \\
HD97658 b  &  -4.975 (17\tablefootmark{a})  &  3.148  &  0.024  &  0.208  &  0.0796  &  5170 / 5137 (H) &  -- / --  &  OK / OK  \\

Kepler-10 b  &  -4.96 (18)  &  3.179  &  0.010  &  0.131  &  0.01685  &  5708 / 5627  &  -- / --  &  OK / OK  \\
Kepler-10 c  &  -4.96 (18)  &  3.484  &  0.054  &  0.209  &  0.241  &  5708 / 5627  &  -- / --  &  OK / OK  \\
Kepler-16(AB) b  &  -4.68 (19)    &  3.162  &  0.333  &  0.753  &  0.7048  &  4450 / 4450  &  -- / --  &  OK / OK  \\
Kepler-17 b  &  -4.47 (20\tablefootmark{a})  &  3.541  &  2.482  &  1.33  &  0.02591  &  5781 / 5781 (H) &  OK / OK  &  OK / OK  \\
Kepler-19 b  &  -4.95 (21)  &  3.607  &  0.064  &  0.198  &  0.085  &  5541 / 5541  &  -- / --  &  OK / OK  \\
Kepler-20 b  &  -4.93 (20\tablefootmark{a})  &  3.358  &  0.026  &  0.170  &  0.04537  &  5466 / 5455  &  -- / --  &  OK / OK  \\
Kepler-20 c  &  -4.93 (20\tablefootmark{a})  &  3.214  &  0.049  &  0.273  &  0.093  &  5466 / 5455  &  -- / --  &  OK / OK  \\
Kepler-20 d  &  -4.93 (20\tablefootmark{a})  &  2.990  &  0.023  &  0.245  &  0.3453  &  5466 / 5455  &  -- / --  &  OK / OK  \\
Kepler-20 e  &  -4.93 (22)  &  3.597  &  0.009  &  0.078 &  0.0507  &  5466 / 5455  &  -- / --  &  OK / OK  \\
Kepler-20 f  &  -4.93 (22)  &  4.139  &  0.045  &  0.09 &  0.11  &  5466 / 5455  &  -- / --  &  OK / OK  \\
Kepler-22 b  &  -5.087 (20)  &  3.791  &  0.11  &  0.21  &  0.849  &  5518 / 5518  &  -- / --  &  OK / OK  \\
Kepler-25 b  &  -5.255 (23)  &  3.107  &  0.030  &  0.241  &  0.068  &  6190 / 6190  &  -- / --  &  OK / OK  \\
Kepler-25 c  &  -5.255 (23)  &  2.950  &  0.077  &  0.463  &  0.11  &  6190 / 6190  &  -- / --  &  OK / OK  \\ 
Kepler-48 b  &  -4.838 (23)  &  3.038  &  0.012  &  0.167  &  0.0532027\tablefootmark{b}  &  5190 / 5190  &  -- / --  &  OK / OK  \\
Kepler-48 c  &  -4.838 (23)  &  3.290  &  0.045  &  0.241  &  0.0851485\tablefootmark{b}  &  5190 / 5190  &  -- / --  &  OK / OK  \\
Kepler-48 d  &  -4.838 (23)  &  3.271  &  0.024  &  0.181  &  0.231\tablefootmark{b}  &  5190 / 5190  &  -- / --  &  OK / OK  \\   
Kepler-62 b  &  -4.863 (24)  &  3.705  &  0.028  &  0.117 &  0.0553  &  4869 / 4925  &  -- / --  &  OK / OK  \\
Kepler-62 c  &  -4.863 (24)  &  4.132  &  0.012  &  0.048 &  0.0929  &  4869 / 4925  &  -- / --  &  OK / OK  \\
Kepler-62 d  &  -4.863 (24)  &  3.557  &  0.044  &  0.174 &  0.12  &  4869 / 4925  &  -- / --  &  OK / OK  \\
Kepler-62 e  &  -4.863 (24)  &  4.131  &  0.113  &  0.144 &  0.427  &  4869 / 4925  &  -- / --  &  OK / OK  \\
Kepler-62 f  &  -4.863 (24)  &  4.235  &  0.11  &  0.126  &  0.718  &  4869 / 4925  &  -- / --  &  OK / OK  \\
Kepler-68 b  &  -5.15 (25)  &  3.148  &  0.023  &  0.205  &  0.0617  &  5793 / 5793  &  -- / --  &  OK / OK  \\
Kepler-68 c  &  -5.15 (25)  &  3.753  &  0.015  &  0.081  &  0.09059  &  5793 / 5793  &  -- / --  &  OK / OK  \\
Kepler-78 b  &  -4.52 (26)  &  3.061  &  0.005  &  0.107  &  0.01  &  5089 / 5089  &  -- / --  &  OK / OK  \\
Kepler-93 b  &  -4.975 (23)  &  3.229  &  0.011  &  0.132  &  0.0542  &  5669 / 5669  &  -- / --  &  OK / OK  \\

Qatar-1 b  &  -4.6 (27)  &  3.300  &  1.095  &  1.164  &  0.02343  &  4861 / 4861  &  OK / OK  &  OK / OK  \\
WASP-4 b  &  -4.865 (28)  &  3.197  &  1.237  &  1.395  &  0.02312  &  5500 / 5513 (H) &  OK / OK  &  OK / OK  \\
WASP-5 b  &  -4.72 (29)  &  3.471  &  1.641  &  1.171  &  0.02729  &  5700 / 5785 (H) &  OK / OK  &  OK / OK  \\                   
WASP-7 b  &  -4.981 (30)  &  3.129  &  0.961  &  1.33  &  0.0617  &  6400 / 6621 (H) &  -- / --  &  -- / --  \\

WASP-15 b  &  -4.86 (29)  &  2.819  &  0.543  &  1.428  &  0.0499  &  6300 / 6573 (H) &  -- / --  &  -- / --  \\
WASP-16 b  &  -5.1 (31)  &  3.319  &  0.857  &  1.008  &  0.0421  &  5550 / 5726 (H) &  OK / OK  &  OK / OK  \\
WASP-22 b  &  -4.9 (32) &  3.036  &  0.588  &  1.158  &  0.04698  &  6000 / 6153 (H) &  OK / OK  &  OK / OK  \\
WASP-23 b  &  -4.68 (11\tablefootmark{a})  &  3.368  &  0.872  &  0.962  &  0.0376  &  5150 / 5046 (H) &  OK / OK  &  OK / OK  \\
WASP-26 b  &  -4.98 (32)  &  3.191  &  1.035  &  1.281  &  0.03985  &  5950 / 6034 (H) &  OK / OK  &  OK / OK  \\
WASP-41 b  &  -4.67 (34) &  3.192  &  0.921  &  1.21  &  0.04  &  5450 / 5546 (H) &  OK / OK  &  OK / OK  \\
WASP-42 b  &  -4.9 (35)  &  3.026  &  0.500  &  1.08  &  0.0458  &  5200\tablefootmark{b} / 5315 (H) &  OK / OK  &  OK / OK  \\
WASP-48 b  &  -5.135 (36)  &  2.940  &  0.994  &  1.67  &  0.03444  &  5990 / 6000  &  OK / OK  &  OK / OK  \\
WASP-50 b  &  -4.67 (37)  &  3.439  &  1.443  &  1.138  &  0.02913  &  5400 / 5518 (H) &  OK / OK  &  OK / OK  \\
WASP-52 b  &  -4.4 (38\tablefootmark{a})  &  2.847  &  0.458  &  1.27  &  0.0272  &  5000 / 5000  &  OK / OK  &  OK / OK  \\
WASP-58 b  &  -4.4 (39\tablefootmark{a})  &  3.071  &  0.892  &  1.37  &  0.0561  &  5800 / 5800  &  OK / OK  &  OK / OK  \\
WASP-59 b  &  -4.1 (38\tablefootmark{a})  &  3.550  &  0.859  &  0.775  &  0.0696866\tablefootmark{b}  &  4650 / 4650  &  OK / OK  &  OK / OK  \\
WASP-69 b  &  -4.54 (39)  &  2.871  &  0.3  &  1.0   &  0.04525\tablefootmark{b}  &  4700\tablefootmark{b} / ~  &  OK / --  &  OK / --  \\
WASP-70 b  &  -5.23 (39)  &  3.366  &  0.6  &  0.8   &  0.04853\tablefootmark{b}  &  5700\tablefootmark{b} / ~  &  OK / --  &  OK / --  \\
WASP-80 b  &  -4.495 (40)  &  3.180  &  0.554  &  0.952  &  0.0346  &  4145 / 4145  &  -- / --  &  -- / --  \\
WASP-84 b  &  -4.43 (39)  &  3.287  &  0.694  &  0.942  &  0.0771  &  5314 / 5314  &  OK / OK  &  OK / OK  \\
WASP-117 b  &  -4.95 (41)  &  2.816  &  0.275  &  1.021  &  0.09459  &  6040 / 6040  &  OK / OK  &  OK / OK  \\

\hline 
\end{longtable}
\tablefoot{The line divides the data in two groups according to their provenience; the top elements are extracted from H10, the bottom ones from the literature. The first  T$_{\mathrm{eff}}$ value for each entry corresponds to the value obtained from the literature and the second from SWEET-Cat, to which an (H) flag follows if it was obtained through an homogeneous analysis. 
\tablefoottext{a}{Data queried using \textit{Exoplanet Orbit Database}.}
\tablefoottext{b}{Data absent from \textit{Exoplanet Orbit Database} and recovered from \textit{Exoplanet Orbit Database}, except for WASp-69b and WASP70b, from \cite{2013arXiv1310.5654A} and the Kepler-48c semi-major axis from \url{http://www.openexoplanetcatalogue.com/system.html?id=Kepler-48+c}.}
}
\tablebib{
(1)~\citet{2010ApJ...717L.138H};  (2)\citet{2004ApJ...614L..81M}; (3)\citet{2009A&A...506..303Q}; (4)\citet{2010ApJ...724..866K}; (5)\citet{2010ApJ...720.1118B}; (6)\citet{2012ApJ...749..134H}; (7)\citet{2012ApJ...745...80Q}; (8)\citet{2011ApJ...728..138H}; (9) \citet{2011ApJ...734..109B}; (10)\citet{2011ApJ...733..116B}; (11)\citet{2011AJ....142...95K}; (12)\citet{2011ApJ...742...59H}; (13)\citet{2012AJ....144...19B}; (14)\citet{2012PASJ...64...97S}; (15)\citet{2012AJ....144..139H}; (16)\citet{2014AJ....147..128H}; (17)\citet{2011ApJ...730...10H}; (18)\citet{2014arXiv1405.7881D}; (19)\citet{2011ApJ...741L...1W}; (20)\citet{2011ApJ...736...19B}; (21)\citet{2011ApJ...743..200B}; (22)\citet{2012ApJ...749...15G}; (23)\citet{2014ApJS..210...20M}; (24)\citet{2013Sci...340..587B}; (25)\citet{2013ApJ...766...40G}; (26)\citet{2013Natur.503..377P}; (27)\citet{2013A&A...554A..28C}; (28)\citet{2011ApJ...727...23B}; (29)\citet{2010A&A...524A..25T}; (30)\citet{2012ApJ...744..189A}; (31)\citet{2012MNRAS.423.1503B}; (32)\citet{2011A&A...534A..16A}; (33)\citet{2011A&A...531A..24T}; (34)\citet{2011PASP..123..547M}; (35)\citet{2012A&A...544A..72L}; (36)\citet{2014ApJ...781..109O}; (37)\citet{2011A&A...533A..88G}; (38)\citet{2013A&A...549A.134H}; (39)\citet{2013arXiv1310.5654A}; (40)\citet{2014A&A...562A.126M}; (41)\citet{2014A&A...568A..81L}.
}\label{TableData}

%\end{landscape}
%\end{longtab}

\end{appendix}

\end{document}